\documentclass{amsart}
\usepackage{amsmath}
\usepackage{latexsym}
\usepackage{amssymb}

\newcommand{\C}{\mathbb{C}}
\newcommand{\Om}{\Omega}
\newcommand{\CA}{\mathcal A}
\newcommand{\sr}{\sigma^{\rm red}}
\newcommand{\CK}{\mathcal K}
\newcommand{\CAr}{{\mathcal A}^{\rm red}}
\newcommand{\Hr}{{H}^{\rm red}}
\newcommand{\Omr}{{\Omega}^{\rm red}}

\newcommand{\CB}{\mathcal B}
\newcommand{\CC}{\mathcal C}
\newcommand{\CE}{\mathcal E}
\newcommand{\CF}{\mathcal F}
\newcommand{\CG}{\mathcal G}
\newcommand{\CH}{\mathcal F}
\newcommand{\CL}{\mathcal L}
\newcommand{\CM}{\mathcal M}
\newcommand{\CO}{\mathcal O}
\newcommand{\cP}{\mathcal P}
\newcommand{\mfg}{\mathfrak g}
\newcommand{\mfk}{\mathfrak k}

\newcommand{\CZ}{\mathcal Z}
\newcommand{\hP}{\hat P}
\newcommand{\hE}{\hat E}
\newcommand{\hX}{{\widehat{X}}}
\newcommand{\tl}{\widetilde{\lambda}}
\newcommand{\bl}{\overline{\lambda}}
\newcommand{\bz}{\overline{z}}
\newcommand{\tm}{\widetilde{\mu}}
\newcommand{\CP}{\mathbb{CP}}
\newcommand{\tX}{\tilde{X}}

\newcommand{\db}{\bar{\partial}}
\renewcommand{\d}{\mathrm{d}}
\newcommand{\rd}{\mathrm{d}}
\newcommand{\p}{\partial}
\newcommand{\bp}{\overline{p}}

\newcommand{\ip}{\frac{i}{2\pi}}
\newcommand{\tr}{\mathrm{tr}}
\newcommand{\rV}{\mathrm{Vect}}
\newcommand{\hV}{\mathrm{HolVect}}
\newcommand{\Det}{\mathrm{Det}}
\newcommand{\Gr}{\mathrm{Gr}}

\newcommand{\half}{{\textstyle \frac{1}{2}}}

\newcommand{\hook}{{\setlength{\unitlength}{10pt}   
                   \begin{picture}(.833,.8)
                   \put(.15,.08){\line(1,0){.35}}
                   \put(.5,.08){\line(0,1){.5}}
                   \end{picture}}}

\newenvironment{remark}{\medbreak \noindent{\bf Remark }}{
\medbreak}

\newtheorem{proposition}{Proposition}

\newcommand{\be}{\begin{equation}}
\newcommand{\ee}{\end{equation}}

\newcommand{\R}{{\mathbb R}}

\calclayout

\begin{document}
\title{Tau-functions, twistor theory, and quantum field theory}
\author{L.J.Mason, M.A.Singer \and N.M.J.Woodhouse}
\address{(Mason and Woodhouse) Mathematical Institute, 
  Oxford, OX1 3LB, UK}
\address{(Singer) Department of Mathematics and Statistics, University
of Edinburgh, King's Buildings, Edinburgh EH9 3JZ, UK}
\email{lmason@maths.ox.ac.uk}
\email{michael@maths.ed.ac.uk}
\email{nwoodh@maths.ox.ac.uk}

\begin{abstract}
  This article is concerned with obtaining the standard tau function
  descriptions of integrable equations (in particular, here the KdV
  and Ernst equations are considered) from the geometry of their
  twistor correspondences.  In particular, we will see that the
  quantum field theoretic formulae for tau functions can be understood
  as arising from geometric quantization of the twistor data.  En
  route we give a geometric quantization formulation of Chern-Simons
  and WZW quantum field theories using the Quillen determinant line
  bundle construction and ingredients from Segal's conformal field
  theory.  The $\tau$-functions are then seen to be amplitudes associated
  with gauge group actions on certain coherent states within these
  theories that can be obtained from the twistor description.
\end{abstract}
\maketitle
\section{Introduction}
One of the most significant overviews of the theory of integrable
systems is that provided by the grassmanian approach of Sato and its
development by the Japanese school into a formulation based on quantum
field theory. The geometric and analytic underpinnings of the
grassmanian approach were further developed in Segal and Wilson
(1985).  This approach was first used to bring out the
(infinite-dimensional) geometry of `equations of KdV type'; the KdV
equation itself as well as $n$-KdV and the KP equation.  The central
construct in this approach is the $\tau$-function which serves as a
`potential' for the dependent variables which appear in the KP
equation (and its specializations $n$-KdV and KdV).  In the paper of
Segal and Wilson, the $\tau$-function is constructed in terms of
infinite determinants.  The Japanese school interpret these as quantum
field theoretic amplitudes of the form
\begin{equation}\label{japfrm}
\tau(x,t)=\langle 0|\exp\{ x\phi_1 + t\phi_2\} |\psi\rangle
\end{equation}
for some state $|\psi\rangle$ and operators $\phi_1$ and $\phi_2$ in a
two-dimensional quantum field theory with vacuum state $|0\rangle$.  Since
this foundational work, $\tau$-functions have been introduced in the
study of many other integrable systems.

Another significant (but more recent) unifying idea in the theory
of integrable systems originates in Richard Ward's observation that
many one and two-dimensional integrable systems are symmetry
reductions of the self-dual Yang-Mills equations. Such systems can be
classified as reductions of the self-dual Yang-Mills equations and
their theory obtained from the complex geometry of twistor theory
which gives, in effect, the general solution of these equations, Mason
\& Woodhouse (1996). In particular twistor methods are applicable to
the study of the KdV and $n$-KdV equations, Mason and Singer
(1994)\footnote{The KP equations do not appear to be a reduction of
  the self-dual Yang-Mills equations with finite dimensional gauge
  group.  However, it is possible even so to find generalized twistor
  correspondences for these equations, Mason (1985), and \S\S12.6 of
  Mason and Woodhouse (1996).}.

This paper is one in a series which is devoted to the clarification of
links between Ward's twistor approach and other pre-existing methods.
Its main purpose is to give a geometric account of the quantum
field-theoretic approach of the Japanese school and its relation to
the geometry of the twistor construction.  It is a sequel to Mason \&
Singer (1994), which focussed on the twistor theory of $n$-KdV
equations and is a parallel development to that in Mason, Singer \&
Woodhouse (2000), which gave a definition of tau-functions as
an infinite dimensional determinant (or cross-ratio).  The purpose of
this paper is to make a more direct contact with quantum field theory
and Quillen determinants.   It has been written so as to be largely
self-contained.

We now give a more detailed outline of the work presented here.  In
\S\S2--4 we give an account of the quantum field theories that are
relevant to KdV and certain other integrable systems.  These theories
are versions of Chern--Simons theory and the WZW model, and have been
much studied [see the cited works by Felder, Gawedzki, Kupiainen,
Gepner and Witten].  However we did not find in the literature a
source which deals with them as presented here. Our treatment seems
very natural; it is an application of the methods of geometric
quantization in an infinite-dimensional setting, combined with
Quillen's determinant line-bundle to give an explicit construction of
the Fock space as the space of holomorphic sections of the prequantum
line-bundle with inner product obtained from Segal's formulation of
conformal field theory using the gluing of determinants.

The ingredients needed for geometric quantization are: first, the {\em
  classical phase space} $\cP$ (a symplectic manifold); second, a
choice of real or complex polarisation; and finally a choice of
prequantum line bundle $\Det \to \cP$, which, as the notation is
intended to suggest, turns out to be Quillen's determinant line
bundle. In this paper, we shall always use a complex polarisation, so
that $\cP$ becomes a K\"ahler manifold.  The bundle $\Det$ is required
to admit a $U_1$-connection whose curvature is $i$ times the
symplectic form of $\cP$.  For Chern--Simons and WZW, these data are
respectively
\begin{itemize}
\item Chern--Simons, \S2--3: $\cP= \CA_+$, the space of unitary
  connections in a trivial bundle over a disc $D_+ \subset \CP^1$; the
  symplectic form is
$$
\Om(a,b) = \frac{1}{2\pi}\int_{D_{+}}\tr(a\wedge b);
$$
$\CA_+$ has a complex structure by identification with the space of
$\db$-operators; and the prequantum line-bundle is essentially
Quillen's determinant bundle.

\item WZW, \S\ref{red}: $\cP=\CAr=$ the space of based loops
 $\Om SU_n       =LSU_n/SU_n$, with  its standard homogeneous
 symplectic form equal to
$$
\Omr(u,v) = \int_{S^1}\tr(u\rd v)\mbox{ at the identity coset.}
$$
The polarisation is given by one of the standard factorization
theorems, which gives $\Om SU_n$ a K\"ahler structure through the
identification with $LSL_n(\C)/L^+SL_n(\C)$. The prequantum
line-bundle is also a determinant bundle and has been constructed in
Segal and Wilson (1985), Pressley and Segal (1986).
\end{itemize}

Given these data, geometric quantization yields a quantum state space
$\CH$ as the space of `square-integrable' holomorphic sections of
$\Det$ over $\cP$. (In the Chern-Simons case we will be concerned with
sections invariant under the action of the group of based gauge
transformations.)  This construction is {\em natural} in that if a
group $\CG$ acts compatibly with the symplectic form and polarisation,
then the action can be quantized (provided a moment map can be found).
If a moment map is obstructed by a cocycle, we obtain a representation
of the central extension of $\CG$ on $\CH$ generated by that cocycle.
We will be interested in the actions both of gauge transformations and
diffeomorphisms of $D_+$.  The moment maps for these symmetries of the
phase spaces can also be obtained from Noether's theorem.

The geometric interpretation of the `Japanese formula' \eqref{japfrm}
has to do with the non-invariance of certain {\em coherent states} in
$\CH$ under symmetries that act holomorphically but not symplectically
on $\cP$ (these will lie in the complexification of a real group of
holomorphic symplectomorphisms).  This leads to a first general
definition of the $\tau$-function in \S\ref{deftau}.  (Another reason
why we are interested in these symmetries is that, as the notation is
intended to suggest, the phase space $\CAr$ is a reduction of the
phase space $\CA_+$ by the group of gauge transformations that are the
identity on $\p D_+$.)

The connection with the more standard quantum field theory notation is
given in \S5.  In this context, the coherent states $|\Psi(p)\rangle$
in $\CH$ correspond to points $p\in \cP$ and $\Psi(p)\in \Det^*_p$.
They arise from the operation of evaluation of holomorphic sections at
$p$.  This defines a linear functional $\CH \to \Det_p$ and then
multiplication by $\Psi(p)$ gives a complex number.  Any such linear
functional is given by pairing with some state of $\CH$---let that
state be denoted $|\Psi(p)\rangle$.  [The simplest example of this
phenomenon is the geometric quantization of the Riemann sphere
$\CP^1$. If we take for the symplectic form $n$ times the area form,
then $\CH$ is the space of polynomials of degree $n$ (the
$n+1$-dimensional irreducible representation of $SU_2$). If $p\in
\CP^1$, then $|p\rangle = (\bp z +1)^n$ up to scale, the polynomial
with an $n$-fold zero at the antipode $-1/\bp$ of $p$.]  The inner
product on the Fock space then can be seen to arise from the Segal
gluing formulae for determinants.

We now explain how these phase spaces arise naturally in the twistor
description of integrable systems such as KdV and the Ernst equations.
The general formulation and details of these two examples appear in
\S\S6--8.  

In the twistor description of integrable systems, the basic
geometric object is a holomorphic vector bundle $\CE$, over an
auxiliary complex manifold $\CZ$ called twistor space. Twistor space
$\CZ$ is related to `space-time' $\CM$ (the space of independent
variables of the integrable system in question) through a {\em
  correspondence} which has the property that the points of space-time
parameterize a family of $\CP^1$'s (so-called twistor lines) in $\CZ$.
Now a $\tau$-function can only be defined from the twistorial point of
view when additional symmetries are imposed upon $\CE$. Technically,
the main requirement is that a group of symmetries should act on $\CZ$
with generic orbit of (complex) codimension $1$. (In the examples
presented here, $\CZ$ has dimension $2$ so we require a
$1$-dimensional symmetry group. Higher-dimensional examples appear in
Mason, Singer and Woodhouse (2000).)  The presence of such symmetries
allows us to pass, in a natural way, from the bundle $\CE$ to a family
of holomorphic structures on a fixed bundle $E$ over $\CP^1$, with an
explicit formula for the variation of holomorphic structure with the
point in space-time.  More
explicitly, we can regard this as a family of $\db$-operators
(parameterized by space-time) on a fixed trivial bundle over $\CP^1$,
or as a similarly parameterized family of patching functions
(clutching functions) in a \v{C}ech description of $\CE$.  The
variation of this holomorphic structure with space-time is 
given by a combination of a (complex) gauge transformation and a
diffeomorphism.  In the KdV case we need only use gauge
transformations, and in the Ernst equation case we need only use a
diffeomorphism. Thus we
have reached a point of contact with the classical phase spaces
described before, for we can regard our family of $\db$-operators as a
finite-dimensional submanifold of $\CA_+$, and our family of patching
functions as a finite-dimensional submanifold of $\CAr$.  Furthermore,
the variation of the $\db$-operators or patching functions in the
family is given by a holomorphic symmetry of the phase space so that
the family is entirely determined by the symmetry once its initial
value is known.  Postponing the details until \S7, the upshot of this
is that we can interpret the right hand side of \eqref{japfrm} within the
geometric quantization framework---$|0\rangle$ and $|\psi\rangle$ are
(suitably normalized) coherent states associated to the initial value
of the trivial solution and of $\psi$ respectively; while the
exponential represents the quantization (representation on $\CH$) of
the translation from $0$ to $(x,t)$.

The construction presented here is based on the geometric definition
of the $\tau$-function given by Segal and Wilson (1985). As discussed
in Mason \& Singer (1994) the twistor description of KdV can be
regarded as a generalization of the description of Segal \& Wilson. It
is a strict generalization because there is a twistor description of
any local (holomorphic) solution of KdV, whereas Segal \& Wilson only
obtain solutions in a certain class (i.e.\ those with a convergent Baker
function---these are, in particular, meromorphic for all complex times).  The
correspondence between the Segal--Wilson description and the twistor
description goes roughly as follows:
\begin{center}\begin{tabular}{cccc}
Segal--Wilson &&& Twistor description \\
$W\subset \Gr$&$\mbox{ }$&$\mbox{ }$& Representation of $\CE$ restricted to a
twistor line \\ 
$\Gamma_+$ &&& Holomorphic symmetry of $\CZ$ \\
\end{tabular}
\end{center}

The main ideas of this paper can be obtained by reading up to
\S\ref{deftau} and then skipping ahead to \S\ref{fock} leaving out any
subsequent material concerning  Cech representations of bundles.

Finally we note that once the formulae of \S3 and \S4 have been
obtained, one could refer to Mason, Singer \&
Woodhouse (2000) for applications to twistorial definitions of
$\tau$-functions. In that paper such formulae were derived from a
slightly different point of view. The reader is also referred to that paper for
the twistorial definition of $\tau$-functions of several integrable
systems not discussed here.

\setcounter{equation}{0}
\section{The space of connections as a classical phase-space}\label{connphase}

In this section we describe in detail the geometry of the space of 
connections over the disc, in particular its symplectic and complex 
structure.  We consider natural groups of symmetries (and algebras of 
infinitesimal symmetries) of this space and the extent to which they 
preserve the symplectic and complex structure. We also discuss the 
reduction of this phase-space to one closely related to loop groups. 
In the next section we shall turn to the problem of quantizing
this phase-space.

Since connections, $\db$-operators and related notions will 
be in constant use throughout this paper, we begin by recalling 
these notions, from a point of view close to that of
Atiyah \& Bott (1982) or Donaldson \& Kronheimer (1990).

\subsection{Connections and $\db$-operators}
\label{condb}
In this section $M$ is a smooth manifold, and $E\to M$ is a complex 
vector bundle of rank $n$, 
with structure group $K$, a (usually compact) Lie group. A 
$K$-connection $A$ on $E$ determines and is determined by the 
covariant derivative operator
\begin{equation}\label{con1}
\nabla_{A}:\Om^{0}(M,E) \longrightarrow \Om^{1}(M,E);
\end{equation}
this is a linear differential operator preserving the $K$-structure 
and satisfying the Leibnitz rule
\begin{equation}\label{lieb}
\nabla_{A}(f\otimes s) = df \otimes s + f\otimes\nabla_{A}s
\end{equation}
for any smooth function $f$ and section $s$ of $E$. (Here 
$\Om^{p}(M,E)$ denotes the space of smooth $p$-forms with values in $E$.) 
The operator in \eqref{con1} extends in a standard way to define a 
covariant exterior derivative
\begin{equation}
    \rd_{A}:\Om^{p}(M,E)\longrightarrow \Om^{p+1}(M,E)
\end{equation}
and this has {\em curvature} $F_{A}= \rd_{A}^{2}$, a $2$-form with
values in the endomorphisms of $E$ that respect the $K$-structure, a
space we shall write as $\Om^{2}(\mfk(E))$.

The space $\CA=\CA(M,E)$ of all $K$-connections on $E$ is an
infinite-dimensional affine space relative to the vector space $\CB=
\CB(M,E)=\Om^{1}(M,\mfk(E))$. In other words the difference of any two
connections in $\CA$ is a $1$-form with values in $\mfk(E)$. Thus the
{\em tangent space} $T_{A}\CA\equiv\CB$, for any point $A$ of $\CA$.
If $a\in \CB$, then the derivative of $F_{A}$ in the direction of $a$
is given by
\begin{equation}\label{dcurv}
\delta_{a}F_{A }= d_{A}a.
\end{equation}
The gauge group $\CK$ of all automorphisms of $E$ respecting the 
$K$-structure acts on $\CA$ by conjugation: if $g\in \CK$, then
\begin{equation}\label{gaac}
    \nabla_{g(A)} = g\cdot \nabla \cdot g^{-1} = \nabla_{A}
    - \nabla_{A}g\,g^{{-1}}.
\end{equation}
Any element $u\in  \Om^{0}(M,\mfk(E))$ determines an infinitesimal 
gauge transformation $g = 1 + \varepsilon u$. Inserting this in 
(\ref{gaac}) and working to first order in $\varepsilon$ we obtain the 
formula $\delta_u\nabla_A = -\rd_A u$. The correct interpretation of 
$\delta_u$ here is as a vector field on $\CA$, whose value at $A$
is $-\rd_{A}u\in \CB$.

Turning now to $\db$-operators, we assume that $M$ is a complex
manifold, so that we can introduce local holomorphic coordinates
$z_{j}$ near any point. Then the space of complex valued $1$-forms
$\Om^{1}$ splits as a direct sum $\Om^{(1,0)}(M)\oplus\Om^{(0,1)}(M)$
generated locally by the $\rd z_{j}$ or the $\rd\bar{z_{j}}$
respectively. (Complex-valued $k$-forms can similarly be decomposed as
$\oplus_{p+q=k}\Omega^{(p,q)}(M)$.) If now $E\to M$ is a complex
vector bundle with complex structure group $K^{c}$, a
($K^{c}$-)$\db$-operator on $E$ is a linear differential operator
\begin{equation}\label{dbdef}
    \Om^{(0,0)}(M,E) \longrightarrow \Om^{(0,1)}(M,E)
\end{equation}
satisfying a Leibnitz rule as in \eqref{lieb} and preserving the
$K^{c}$ structure of $E$. (Later on, we shall only need $K^c =
SL_n(\C)$ so that all $\db$-operators are required to annihilate a
holomorphic $n$-form on $E$.) Extending $\db_{\alpha}$ to act on
$\Om^{(0,p)}(M,E)$, we introduce the algebraic operator
$\db_{\alpha}^{2}$, a $(0,2)$-form with values in $\mfk^{c}(E)$. We
say that $\db_{\alpha}$ is integrable if $\db_{\alpha}^{2}=0$. Any
integrable $\db$-operator defines a holomorphic structure on $E$; the
local holomorphic sections are those that are annihilated by
$\db_{\alpha}$ and there exist enough such local sections to form
holomorphic frames near any point of $M$.


In general, the space $\CA^{c}=\CA^{c}(M,E)$ of $K^{c}$--$\db$-operators on 
$E$ is an infinite-dimensional complex affine space relative to the vector 
space $\CB^{c}=\CB^{c}(M,E)= \Om^{(0,1)}(M,\mfk(E))$. The group of 
complex gauge transformations $\CK^{c}$ acts on $\CA^{c}$, analogously 
to \eqref{gaac}:
\begin{equation}\label{cggg}
    \db_{g(A)}= g\cdot\db_{\alpha}\cdot g^{{-1}} = \db_{\alpha}- 
    \db_{\alpha}g\,g^{{-1}};
\end{equation}
This preserves integrability, and indeed two integrable 
$\db$-operators define isomorphic holomorphic structures on $E$ iff 
they are (complex)-gauge equivalent. If $u$ is an infinitesimal 
complex gauge transformation, then it defines a {\em holomorphic} 
tangent vector field $\delta_{u}= -\db_{\alpha}u$ on $\CA^{c}$.

If one fixes a choice of a hermitian structure, a $\db$--operator
gives rise to a unitary connection.  This follows from Chern's
theorem: there is a unique unitary connection whose $(0,1)$ part
defines any given holomorphic structure.  When $M$ has complex
dimension $1$ the integrability condition is trivially satisfied so
that on a bundle with fixed hermitian structure $\db$-operators are in
1--1 correspondence with unitary connections.  It follows that if
$K = U(n)$ or $SU(n)$, then $\CA(M,E)=\CA^{c}(M,E)$. 
This identification simply maps $\rd_A$ to its 
$(0,1)$-part $\rd_A^{(0,1)}= \db_A$  and Chern's result asserts that 
this map is an isomorphism.  Relative to a unitary
trivialization such that $\rd_A = \rd + A$, we write $A= \alpha -
\alpha^*$, where $\alpha\in\CB^c$, so that $\db_A = \db +\alpha$,
$\p_A = \p - \alpha^*$.  

An important corollary of this identification between the space of
$\db$-operators with the space of connections is that it leads to a
natural action of the group of complex gauge transformations on $\CA$:
if $g$ is such a complex gauge transformation, its action is given by
$$
\rd_{g(A)} = g\cdot\db_A\cdot g^{-1} + (g^*)^{-1}\cdot \p_A \cdot g^*.
$$
We now study the symplectic geometry of this space.

\subsection{Connections on domains in $\C$}

We will be interested in the space of connections on certain domains
with boundary in $\C$.  Let $D_{-}$ be a finite disjoint union of open
discs in $\CP^{1}$ and let $D_{+}$ be the complement of $D_{-}$. Then
$D_{+}$ is a closed subset of $\CP_{1}$ with non-empty interior
$D_{+}^{0}$. We think of $D_{-}$ as being a neighbourhood of certain
points `at $\infty$' in $\CP^{1}$. Let $E_{+}\to D_{+}$ be the trivial
complex vector bundle of rank $r$ and with structure group $SU(r)$;
let $\CA_{+}$ be the space of $SU(r)$-connections or equivalently
$SL(r,\C)$-$\db$-operators that are $C^{\infty}$ up to the boundary of
$D_{+}$.

We can think of $\CA_+$ as an infinite-dimensional classical phase
space, for it carries a natural symplectic form $\Om$, given by
$$
\Om(a,b) = \frac{1}{2\pi}\int_{D_{+}}\tr(a\wedge b)\mbox{ for }a,b\in\CB_+
$$
(The normalization factor of $2\pi$ will be convenient later.)  In
this language the complex structure on
$\CA_+$ is a (positive) complex polarization; in other words, $\CA_+$
is an infinite-dimensional K\"ahler manifold.

\subsubsection{The action of gauge transformations on $\CA_+$}
Because $D_+$ is a manifold with boundary, it is natural to
distinguish inside the group $\CG^c$ of complex gauge transformations
that are $C^\infty$ up to the boundary, the normal subgroup $\CG_0^c$
of {\em based} gauge transformations: those that are equal to the
identity on the boundary. Similarly we denote by $\CG \subset \CG^c$
and $\CG_0\subset\CG_0^c$ the subgroups of unitary gauge
transformations.  We shall denote by $\mfg_0,\mfg_0^c,\mfg, \mfg^c$
the Lie algebras of $\CG_0,\CG_0^c,\CG,\CG^c$.

 With regard to the action of these groups on $\CA_+$,
we have the following 
\begin{proposition}
\begin{itemize}
\item The actions of $\CG_0$ and $\CG$ on $\CA_+$ preserve $\Om$ and the
complex polarization;

\item The actions of $\CG_0^c$ and $\CG^c$ on $\CA_+$ preserve the complex
polarization but not $\Om$.

\item If $u\in \mfg$, then a Hamiltonian for $u$ is given by
\begin{equation} \label{hamf1}
H_{u}(A) = -\frac{1}{2\pi}\int \tr(F_{A}u) + \frac{1}{2\pi}\oint\tr(Au). 
\end{equation}

\item The map $u \mapsto H_u(A)$ is not a co-momentum map for $\mfg$ since
$$
\{H_u,H_v\}-H_{[u,v]}= \frac{1}{2\pi}\oint \tr(u\,\rd v).
$$
Instead $u \mapsto H_u(A)$ is a co-momentum map from the central
extension $\widetilde{\mfg}$ of $\mfg$ into $C^\infty(\CA_+)$ with 
cocycle 
\begin{equation}\label{ccy}
c(u,v) =\frac{1}{2\pi}\oint \tr(u\,\rd v).
\end{equation}
However, the cocycle vanishes on $\mfg_0$ and $u\mapsto H_u(A)$
is a moment map $\mfg_0 \to C^\infty(\CA_+)$.
\end{itemize}
\end{proposition}

\begin{remark} In \eqref{hamf1}, $\int$ denotes an integral over
$D_+$, $\oint$ denotes an integral over $\p D_+$. In the
boundary-term, $A$ appears, which is not gauge-invariant. This
indicates that the most natural framework is to use bundles over
$D_+$ that are framed over the boundary. This
fits in naturally with our requirements in subsequent sections, for we
shall want to extend our $\db$-operators in a standard way to operate
on bundles over $\CP^1$. The reader may alternatively take the
view that we are working on the product bundle
over $D_+$ which provides a preferred gauge in which to write $A$. 
\end{remark}
\begin{proof} It is clear that $\CG^c$ preserves the complex
polarisation of $\CA_+$. To verify that $\CG$ preserves the symplectic
structure, it is enough to verify that \eqref{hamf1} is indeed a
Hamiltonian. Without the boundary term, this calculation is in
Atiyah--Bott (1982). 

To verify that $H_{u}$ is a Hamiltonian for the vector field $d_{A}u$ 
on $\CA_{+}$. Let $a$ represent a variation in $A$. Then we have 
$\delta_{a}A=a, \delta_{a}F_{A} = d_{A}a$ and so
$$
\delta_{a}H_{u}(A)= - \frac{1}{2\pi}\int\tr(\rd_{A}a\,u)+\frac{1}{2\pi}\oint 
\tr(a\,u).
$$
Integrating by parts in the first term, using
$\tr(\rd_{A}a\,u)=\rd\,\tr(a\,u) - \tr(a\wedge \rd_{A}u) $, we obtain
$$
\delta_{a}H_{u}(A)= \Omega(a,\rd_{A}u)
$$
as required. Next we compare the
Poisson bracket of $H_{u}$ and $H_{v}$, for two elements $u,v$ of 
$\mfg$, with $H_{[u,v]}$. The Poisson bracket is equal to the 
variation of $\delta_{u}H_{v}$, where we have written $\delta_{u}$ 
for $\delta_{\rd_{A}u}$. From above, this is simply
$$
\delta_{u}H_{v}- H_{[u,v]} 
= \frac{1}{2\pi}\int\tr(\rd_{A}u\wedge\rd_Av)
+\frac{1}{2\pi}\int\tr(F_{A}[u,v])-\frac{1}{2\pi}\oint\tr(A[u,v])
= 
\frac{1}{2\pi}\oint\tr(u\,\rd v)
$$
using the definition of curvature, $\rd_{A}^{2}u = [F_{A},u]$ and 
writing $\rd_{A}u = \rd u + [A,u]$ in the boundary term.
\end{proof}
The subgroups of based gauge transformations play a different role
from the full groups of gauge transformations.  The unbased
transformations will generate the dynamics of our system when we consider
the quantization, whereas the subgroups of based gauge transformations
will corresponds to `genuine' gauge degrees of freedom. All of (the
central extension of)
$\CG^{c}$ acts on the quantum Hilbert space but only $\CG^{c}_{0}$
preserves the `vacuum state'.

\subsubsection{The action of diffeomorphisms on $\CA_+$}
We turn now to consider the action of diffeomorphisms on $\CA_{+}$. 
The most obvious group that acts consists of diffeomorphisms of $D_{+}$ 
that are smooth up to the boundary and tangent to it. Working at the 
infinitesimal level, we introduce the Lie algebra of this,
$\rV_{0}(D_{+})$ and  its complexification $\rV_{0}^{c}(D_{+})$. We
shall also need to consider the algebra
$\hV(D_{+})$ of real vector fields whose $(1,0)$ part is holomorphic in the 
interior of $D_{+}$.  
If $\xi\in \rV_{0}(D_{+})$, then the action on $\CA_{+}$ is 
given by Lie-derivative; the vector field at $A\in \CA_{+}$ is given 
by
$$
\CL_{\xi}(A) = d(A(\xi)) + \xi\hook dA = d_{A}A(\xi) +\xi\hook F_{A}.
$$
Here we have written $A(\xi)$ for the interior product $\xi\hook A$
and the second formula follows from the first by adding and
subtracting the term $\xi\hook A\wedge A$. The fact that $A$ appears
explicitly here reflects the need to make a choice of `invariant'
trivialization of $E$ when lifting the action of the diffeomorphism
group to $\CA$.

Now one can check that a Hamiltonian for this action is given by
\begin{equation} \label{hamf2}
H_{\xi}(A)= \frac{1}{2\pi}\int\tr(F_{A}\,A(\xi))-\frac{1}{4\pi}\oint 
\tr(A\,A(\xi)).
\end{equation}
The verification is easiest using the second formula for 
$\CL_{\xi}(A)$, and also requires the identity $F_{A}\,a(\xi) = 
a\wedge(\xi\hook F_{A})$ which is valid in 2 dimensions. Using these,
one verifies that
$$\delta_{a}H_{\xi}(A)= \Om(a,\CL_{\xi}A).
$$
 Hence 
$\rV_{0}(D_{+})$ acts symplectically on $\CA_{+}$ and in this case there 
is no cocycle.  Writing $A=\alpha d\bar{z}-\alpha^{*}dz$ and $\xi = 
\xi^{1,0}\p_{z}+\xi^{0,1}\db_{z}$,  we have
$$
\CL_{\xi}(A) = [\db_{z}(\xi^{0,1}\alpha) + \xi^{1,0}\p_{z}\alpha - 
\alpha^{*}\db_{z}\xi^{1,0}]d\bar{z} - \mbox{hermitian conjugate}
$$
so that the complex polarisation is not preserved unless
$\db_{z}\xi^{1,0}=0$. Hence the subalgebra $\hV(D_{+})$ acts
preserving the polarisation (but not, in general, the symplectic
structure as such vector fields are generically not tangent to the
boundary). To summarize:
\begin{proposition} The algebra $\rV(D_+)$ of vector fields on $D_+$
acts symplectically on $\CA_+$. The map $\rV(D_+)\to C^\infty(\CA_+)$
given by $\xi \mapsto H_\xi$, with $H_\xi$ given by \eqref{hamf2} is
an equivariant moment map. $\rV(D_+)$ does not preserve the complex
polarisation of $\CA_+$. The algebra $\hV(D_+)$ of vector
fields that are smooth up to the boundary of $D_+$ and whose $(1,0)$
part is holomorphic in
the interior, acts on $\CA_+$ preserving its complex polarisation (but
not, in general, the symplectic form).
\end{proposition}

\noindent{\bf Remark:} The above phase space is that for Chern-Simons
theory, with action
$$
S[A]=\int\tr\left( A\wedge\d A+\frac{2}{3}A^3\right )
$$
and one can verify, by use of Noether's theorem etc., that the above
symplectic form and Hamiltonians arise from this action.
 
\setcounter{equation}{0}
\section{Geometric quantization of $\CA_+$}
\label{gq}
Given a classical phase-space, the {\em pre-quantum data} consist of a
complex line bundle equipped with metric and compatible connection
$\nabla$, such that the curvature of $\nabla$ is equal to a fixed
multiple of 
the symplectic form.  Given also a polarisation, the quantum
phase space is defined to be the vector space of sections of the
pre-quantum line bundle that are flat along the leaves of the
polarisation. In the case of a complex polarisation, this forces one
to look at the holomorphic sections.  Our task in this section is to
quantize $\CA_+$ in this sense and to consider the extent
to which the `classical symmetries' of $\CA_+$ can be implemented as
symmetries of the quantum phase space.

In order to determine pre-quantum data, one can appeal to general
theorems asserting their existence under appropriate circumstances. We
prefer, however, to define these data explicitly using Quillen's
construction of the determinant line-bundle of a family of
$\db$-operators. This theory applies in the first instance to the
space of $\db$-operators over a compact Riemann surface. We shall
reduce to this case by extending $\db_\alpha \in \CA_+$ in a standard
fashion to give a discontinuous $\db$-operator on a bundle over
$\CP^1$. This is essentially equivalent to the imposition of
Atiyah-Patodi-Singer type boundary conditions on $\db_\alpha$ as an
operator over $D_+$. The small price we pay for this is the
discontinuity in the $\db$-operator over $\CP_1$. We 
explain in \S\ref{ujmp} why this does not cause any major difficulties.

First, however, we shall give a brief review of Quillen's
construction, and describe in particular how it simplifies when the
base is $\CP^1$. The last part of this section is concerned with the
implementation of classical symmetries on the quantum phase spaces and
contains the formulae needed for our subsequent definition of the
$\tau$-function.

\subsection{Quillen's determinant construction}

In this section we first review Quillen's construction of the
determinant line-bundle over the space of $\db$-operators over a
compact complex $1$-dimensional manifold (Riemann surface) $M$,
explaining in particular how the 
construction simplifies when $M = \CP^1$.

Let $E \to M$ be a smooth complex vector bundle of rank $n$ over a
compact Riemann surface $M$. As before, $\CA$ is the space of all
$\db$-operators on $E$. We assume that the generic element of $\CA$ is
invertible (as a map $\Om^{0,0}(M,E)\to\Om^{0,1}(M,E)\,$), equivalently
that the index of any element is $0$. By Riemann--Roch this 
condition is just the constraint ${\rm deg}(E) = n({\rm genus}(M)-1)$.

Quillen shows how to define a holomorphic line-bundle $\Det\rightarrow \CA$
with a canonical holomorphic section $\sigma$, also denoted
$\det$. Intuitively, $\sigma(A)$ is the determinant 
of $\db_A$; in particular, $\sigma(A) \not= 0$ iff $\db_A$ is
invertible and so is non-vanishing at generic points of $\CA$.

The next step is to define a hermitian metric on $\Det$; then by
Chern's theorem $\Det$ will acquire a unique unitary connection
$\nabla$ compatible with the holomorphic structure. For this,
additional choices must be made. Quillen picks hermitian metrics on
$E$ and $M$ and defines a hermitian metric on $\Det$ by
$\zeta$-function regularization. He gives a formula for $\nabla
\sigma$ (reproduced below) and proves that the curvature of $\nabla$
is the standard symplectic form $\Om$ on $\CA$:
\begin{equation}\label{symp}
\ip \Omega(\theta^*,\theta)=\ip\int \tr(\theta^*\wedge \theta) 
\end{equation}
where $\theta \in \CB^c = \Om^{(0,1)}(M,\mfk(E))$ represents a 
$(1,0)$-tangent vector to $\CA$. 
In terms of the local formulae
\begin{equation} \label{rk1}
\db_A = d\bar{z}(\partial_{\bar{z}} + \alpha)\;\;\;
\partial_A = dz(\partial_{z} - \alpha^*),
\end{equation}
Quillen gives the following formula 
for the covariant derivative of $\sigma$ in the direction $\theta$:
\begin{equation}\label{derivfrm}
\nabla_\theta\sigma(A) = \sigma(A) \int \tr(J_A \wedge\theta) 
\quad \mbox{ where } \quad 
J_A = \ip dz(\beta - \alpha^* - \frac{1}{2}\partial_z \log \rho),
\end{equation}
this formula being valid at all points $A$ at which $\sigma(A)\not=0$.
In \eqref{derivfrm},
$\alpha^*$ is as before, $\rho$ is the local conformal factor
for the metric $ds^2=\rho^2 |dz|^2$ on $M$ and $\beta dz$ is
a globally defined 1-form which arises from the expansion near the
diagonal of the Schwartz kernel $G_A(z,z')$ of $\db_A^{-1}$
$$
G_A(z,z') = \ip\frac{dz'}{z-z'}(1 + (z-z')\beta(z') -
(\bar{z}-\bar{z}')\alpha(z') + \ldots).
$$

\subsection{Calculation when $M = \CP^1$}

We now take $M=\CP^1$ and $E = \C^n\otimes H^{-1}$ where $H^{-1}\to
\CP^1$ is the tautological bundle $\C^2\mapsto \CP^1$ (dual to the
hyperplane bundle $H$) and is the unique line bundle of degree $-1$.
Let $\CA$ be the space of $\db$-operators on $E$. Any element of $\CA$
has index zero and the generic element is invertible. Note that in
this case the space $\CB$ is equal to $\Omega^{0,1}(\CP^1,{\rm
  End(E)})$ as the twist by $H^{-1}$ cancels.

By Grothendieck's theorem, if $\db_A$ is such an invertible element,
there exists a gauge transformation $g$ such that 
\begin{equation}\label{gg}\db_A = g\db_0
g^{-1},
\end{equation}
 where $\db_0$ is the standard $\db$-operator on
$\C^n\otimes\CO(-1)$. The main purpose of this section is to use $g$
to simplify Quillen's formulae.

Now $\db_0^{-1}$ has  
Schwartz kernel given by 
$$
G_0 = \ip \frac{\rd z'}{z-z'}.
$$
We have written this in terms of local coordinates $(z,z')\in
\CP^1\times \CP^1$. However, $G_0$ extends canonically to define
a smooth section of
$\mbox{pr}_1^*E \otimes \mbox{pr}_2^*[
\Lambda^{1,0}\otimes E^*]$ over $\CP^1\times \CP^1 -\Delta$ and hence
defines canonically an operator 
$$
\Omega^{0,1}(\CP^1,E) \to \Omega^{0,0}(\CP^1,E).
$$
Using the gauge transformation in \eqref{gg}, we have that the
Schwartz kernel of $\db_A^{-1}$ is just $G_A(z,z')
= g(z) \circ G_0(z,z') \circ g(z')^{-1}$  and expanding near the
diagonal we find 
\begin{equation}\label{bg}
\beta dz = \partial g g^{-1}.
\end{equation}
In what follows we shall restrict to the subspace $\CA_0$ of operators
which differ from $\db_0$ by trace-free elements $\CB_0$ of $\CB$. In that
case we can assume that $g$ in \eqref{gg} has unit
determinant so
that $\beta$ in \eqref{bg} is also trace-free.

The Quillen connection requires choices of
hermitian structures on $E$ and $\CP^1$ (although the final
$\tau$-function formulae will be independent of them). With such
choices, the unitary connection corresponding to $\db_A$ is given by
$\rd_A=\db_A+\p_A$ where
$$
\partial_A = (g^*)^{-1}\circ \partial_0 \circ g^*
$$
and we obtain the formula
$$
\rd z(\beta - \alpha^*) = \partial g g^{-1} + g^{-*}\partial g^* =
g(h^{-1}\partial h)g^{-1}
$$
where $h = g^*g$. Recall that ${\rm End}(E)$ is canonically isomorphic
to the bundle of endomorphisms of the trivial rank-$r$ bundle over
$P_1$. Furthermore, the induced action of $\db_0$ on ${\rm End(E)}$
coincides with that of the standard $\db$-operator on the trivial
bundle.  Thus we drop the distinction between $\db_0$ and $\db$
when acting on $g$, $g^*$ or $h$.)

Substituting into \eqref{derivfrm} we obtain the basic formula
\begin{equation} \label{rk2}
\nabla_\theta\sigma = \left(\ip\int\tr(h^{-1}\partial h \wedge
(g^{-1}\theta g))\right)\cdot\sigma 
\end{equation}
for any $\theta\in \CB_0$. Notice that the term in $\rho$ disappears
because $\theta$ can be assumed to be trace-free.
Other useful formulae are 
\begin{equation} \label{rk3}
\rd_A = g\circ\db_0\circ g^{-1} + g^{-*}\circ\partial_0\circ g^* =
g\circ(\db_0 + \partial_0 
+ h^{-1}\partial h)\circ g^{-1}
\end{equation}
so that
\begin{equation} \label{rk35}
F_A = gF_0g^{-1} + g\db(h^{-1}\partial h)g^{-1}
\end{equation}
where 
\begin{equation} \label{rk4}
F_A = \db_A\partial_A + \partial_A\db_A
\end{equation}
is the curvature 2-form.
By our choice of metric on $E$, $F_0$ is multiple of the
identity. Hence \eqref{rk35} can also be written:
\begin{equation} \label{rk5}
g^{-1}(F_A-F_0) g = \db(h^{-1}\partial h).
\end{equation}

\subsection{$\CA_+$ as a space of connections with jumps on $\CP^1$}
\label{ujmp}
Consider now $\CP^1 = D_+\cup D_-$ and the standard bundle $E = 
\C^n\otimes H^{-1}$ over $\CP^1$.  Denote by $E_{\pm}$ the 
restrictions of $E$ to $D_{\pm}$. We may identify $E_+$ with the 
trivial bundle over $D_+$ in such a way that $\rd_0= \rd$. (We choose 
the metrics on $H$ and $\C^n$ to restrict to constant metrics over 
$D_+$.)  Now given any element $A\in \CA_+$, we may regard it as a 
connection on $E_+$ and extend it by $\rd_0$ over $D_-$ to the whole 
of $E$. We shall denote this `extension by zero' of $\rd_A$ also by 
$\rd_A$; it is a connection on $E$ with a simple jump discontinuity 
across $\p D_+$. It will be seen later that such
operators are required for the simplest formulation of various
important ingredients such as the Fock space
inner product etc.

We shall assume in what follows that Quillen's construction extends to
the $\db$-operators with jump discontinuities that result by taking
the $(0,1)$-part of the extension by zero of $\rd_A$ and in particular
that \eqref{derivfrm} continues to hold for such operators.  As a
partial justification for this, observe first that if $\db_\alpha$ is
a $\db$-operator on $E$ with jump discontinuity at $\p D_+$, then
there is a continuous complex gauge transformation $c$, say, of $E$
such that $c\cdot\db_\alpha\cdot c^{-1}$ is smooth. A sketch of the
proof of this is as follows. If we can find, near any point $p$ of $\p
D_+$ a continuous matrix-valued function $u$ which solves the equation
$$
\db_z(1+ u) + \alpha(1+u) = 0 \quad \mbox{ near } p
$$
(in the sense of distributions) then the required gauge transformation
$c$ can be obtained by patching such solutions together by a partition
of unity.  However the usual proof (cf.\ for example Donaldson and
Kronheimer (1990), Chapter 2) yields such a $u$ that is locally in
$L^q$, for any $q>2$.  Since then $\alpha u$ is also in $L^q$, the
ellipticity of $\db$ ensures that $u$ is actually in $L^q_1$ near
$p$. By the Sobolev embedding theorem in 2 dimensions, such a $u$ is
continuous if $q > 2$.

In particular, the discontinuous $\db$-operator $\db_\alpha$ is
invertible, as an operator between appropriate Sobolev spaces, if and
only if there exists a continuous (complex) gauge transformation $g$
of $E$ satisfying \eqref{gg}. It follows that the formula
\eqref{derivfrm} makes sense in this case, the integrand being bounded
on $\CP^1$. (The 1-form $\beta$ is smooth away from $\p D_+$, where it
has at worst a jump discontinuity.)  In order that $\theta$ represent
a tangent vector to $\CA_+$ we must take $\theta$ to be the extension
by zero to $D_-$ of a 1-form that is smooth in $D_+$. Then the
integral in \eqref{derivfrm} extends only over $D_+$.

It is perhaps worth pointing out that even if $A$ vanishes near $\p 
D_+$, so that its extension to $\CP^1$ is actually smooth, the gauge 
transformation of \eqref{gg} does not in general vanish on $D_-$. In 
particular the boundary integrals that we shall see below will not 
generally vanish even in this case.  

This formulation is equivalent to considering a family of
$\db$-operators on a Riemann surface with boundary, using
Atiyah--Patodi--Singer boundary conditions to make such operators
Fredholm.  It would be interesting to consider the analytic issues
involved in giving a more systematic derivation of the connection on a
determinant line-bundle for this case.  However, from now on, we take
over \eqref{derivfrm} to calculate the covariant derivative of the
section $\sigma$ of $\Det$ over $\CA_+$ and use this formula without
further comment.

\subsection{Quantization and  the action of gauge transformations and
vector fields}

Since $\Det$ is a holomorphic line bundle with connection whose
curvature is the symplectic form, it is the prequantum line bundle.
To obtain the quantum Hilbert space one must introduce a polarisation
on $\CA_+$ and consider `polarised' sections of $\Det$.  In this case
the polarisation is the complex structure, and so the space of
holomorphic sections of $\Det$ will yield the quantum Hilbert space
associated to the classical phase space $\CA_+$.  In the following we
wish to lift the action on $\CA_+$ of gauge transformations and
diffeomorphisms of $D_+$ to act on holomorphic sections of $\Det$.  In
the case of gauge transformations, we will see that the action is
immediately holomorphic and so acts directly on the quantum Hilbert
space.  However, in the case of diffeomorphisms, only a subalgebra of
vector fields acts holomorphically.

\subsubsection{Lifting the action of gauge transformations}
First we follow the standard recipe from geometric quantization to lift
the action of the Lie algebra ${\mathfrak g}$ of infinitesimal gauge
transformations to $\Det$.  If $u \in \CG$, the corresponding vector
field on $\CA_{+}$ is $\delta_u=-d_{A}u$. The geometric quantization
lift is 
\begin{equation}\label{gqrec}
\CL_u = \nabla_u -i H_u
\end{equation}
where we have written $\nabla_{u} = \nabla_{\delta_{u}}$ in order to 
simplify the notation.
\begin{proposition}\label{lifg} The recipe \eqref{gqrec} gives the formula
\begin{equation} \label{basic1}
    \frac{\CL_{u}\sigma}{\sigma}=
    -\frac{1}{2\pi i}\oint\tr(\rd g \,g^{{-1}}u).
\end{equation}
\end{proposition}
\begin{proof}
    From \eqref{rk2} and \eqref{gg},
    $$
  \frac{  \nabla_{u}\sigma}{\sigma} = \frac{1}{2\pi i}
    \int\tr(h^{-1}\p h\wedge g^{-1}\db_{A}u\,g)
    = \frac{1}{2\pi i}\int\tr(h^{-1}\p h\wedge \db_0(g^{-1}ug)).
    $$
In order to integrate by parts, note
$$
\rd\,\tr(h^{-1}\p h\,g^{-1}u g) = \tr(\db(h^{-1}\p h)g^{-1}ug)
               -\tr(h^{-1}\p h\wedge\db(g^{-1}ug)).
               $$
But from \eqref{rk5} and the fact that $u$ is trace-free, the first term 
on the right hand side is equal to $\tr(F_{A}u)$, one of the terms in the 
Hamiltonian. On the left hand side use \eqref{rk3} to write
$$
gh^{-1}\p h\,g^{-1} = \rd_{A}-g\cdot\rd\cdot g^{-1}
= \rd_{A}- \rd_{0} - \rd_{0}g\,g^{-1}.
$$
Hence we find
$$
\frac{\nabla_{u}\sigma}{\sigma}= iH_{u}(A) -\frac{1}{2\pi i}\oint\tr(\rd 
g\,g^{-1}u).$$
\end{proof}

\begin{remark}
  (1) Assuming that the connection on $\Det\to \CA_{+}$ really is
  globally defined, this formula is also globally defined since both
  the connection and Hamiltonian are.
  \\
  (2) It is clear from the formula that this action of
  $\widetilde{\mfg}$ on $\Det$ is holomorphic, for its action upon
  $\sigma$ is multiplication by a function that is holomorphic on the
  dense open set of $\CA_{+}$ where $\sigma\not=0$.  It follows that
  one obtains a holomorphic action of $\mfg^{c}$ also, simply by
  replacing $u$ by a complex element of $\mfg^{c}$.
  \\
  (3) In this set-up we have worked with bundles framed over $D_{-}$.
  In the parallel development of this work, Mason, Singer \& Woodhouse (2000),
  this framing is viewed as coming from another solution.  We shall
  not pursue that viewpoint in this paper.
\end{remark}

\subsection{The first definition of the $\tau$-function}\label{deftau}
The framed subalgebra $\mfg_{0}^{c}$ clearly preserves $\sigma$.  We
will see that the natural interpretation is of $\sigma$ as a `vacuum
state'. The framed gauge transformations, being the true degrees of
gauge freedom, fix the vacuum state, while the unframed gauge
transformations shift it. The $\tau$-function in its most general form
is a function on an orbit in the phase space under the action of some
submanifold of $\CG^{c}$.  It is the value of $\sigma$ on that orbit
expressed in an invariant frame of $\Det$.  Later we will give a
quantum field theoretic formulation in which it measures the amplitude
of the two vacuum states related by a complex gauge transformation.

More precisely, given a submanifold (usually subgroup) of
$\CG^{c}$ 
parametrized by $t\mapsto G(t)$, and an initial
connection $\alpha$, define
\begin{equation}\label{deftau0}
\tau(t)=\widehat{G(t)}{}^{-1}\sigma(G(t)\alpha)
\end{equation}
where $\widehat{G(t)}$ is the action of $G(t)$ lifted to $\Det$.  This
only defines $\tau$ as a function up to an overall constant as it is
an element of the fibre of $\Det$ at $\db_{g(0)\alpha}$ which is not
canonically trivial.  Furthermore, $\tau$ is not well defined in
general since the action of $\CG^{c}$ is generally projective, and so
we must require that the submanifold of $\CG^c$ must be one on which
the central extension \eqref{ccy} splits.

This can be seen more clearly in the infinitesimal version of this
definition.  Differentiation of \eqref{deftau0} leads to a 1-form
`$\rd \log\tau$' on $\widetilde{\CG}^{c}$ given by the formula
\begin{equation}\label{deftau1}
\CL_{u}\sigma = (u\hook\rd\log\tau)\sigma
\end{equation}
where $u=g^{-1}\p_t g$.  The 1-form $\rd \log \tau$ always exists, and
when it is closed on restriction to some submanifold of
$\widetilde{\CG}^{c}$ it defines $\tau$ on that submanifold up to a
constant.  However,  in spite of the notation, $\rd\log \tau$ is not
generally closed, instead we have 
$$
v\hook u\hook \rd (\rd \log \tau)= c(u,v) 
$$
where $c(u,v)$ is the cocycle  \eqref{ccy} for the central extension,
and so the submanifold of $\CG^{c}$ must be one on which the cocycle vanishes.

This is sufficient for the definition of the $\tau$-function of the
KdV equation and the reader may wish to skip ahead to section
\S\ref{s2} for this.  In the remainder of this section, we treat the
quantization of diffeomorphisms and in the next section we give a
similar treatment of the quantization of the reduced phase space
$\CAr=\CA_+|_{F_A=0}/\CG_0$.

\subsection{Lifting the action of vector fields}
For the action of the algebra of vector fields $\rV_{0}(D_{+})$, we 
substitute $X = \CL_{\xi}A$ in \eqref{rk2},
\begin{equation} \label{horr}
\frac{2\pi i\nabla_{X}\sigma}{\sigma} = - \int\tr(h^{-1}\p h\wedge\db(
g^{-1}A(\xi)g)) - \int\tr(h^{-1}\p h\wedge \xi\hook\db(h^{-1}\p h))
\end{equation}
where we have rearranged the first term as in the proof of
Proposition~\ref{lifg}.
In order to integrate by parts 
in the first term, note the identity
$$
\rd\,\tr(h^{-1}\p h g^{-1}A(\xi) g) = \tr(F_{A}A(\xi)) - \tr(h^{-1}\p
h\wedge\db(g^{-1}A(\xi)g)).
$$
To integrate the second term in \eqref{horr} by parts, note
$$
\rd \left(\frac{1}{2}\tr(h^{-1}h_{z})^{2}\xi^{(1,0)}\,\rd z \right)
= \tr(h^{-1}h_{z}\db(h^{-1}h_{z}))\xi^{(1,0)}\wedge\rd z\, + 
\frac{1}{2}\tr(h^{-1}h_{z})^{2}\db\xi^{(1,0)}\wedge\rd z
$$
where the first term on the right hand side is equal to the second term on the right hand side 
of \eqref{horr}. Combining these with the formula \eqref{hamf2}, we 
obtain
\begin{multline}\label{horr1}
 \frac{\nabla_{X}\sigma\,}{\sigma} -i H_{\xi}(A) =
    \frac{1}{4\pi i}\int\tr(h^{-1}h_{z})^{2}\db\xi^{(1,0)}\wedge \rd z
    \\ +\frac{1}{4\pi i}\oint\tr[2gh^{-1}\p h\,g^{-1}A(\xi) - A\,A(\xi) - 
    (h^{-1}h_{z})^{2}\xi^{(1,0)}]\rd z.
\end{multline}
This is simplified, using the fact that, with $\xi$ tangent to the
boundary, $\xi^{(1,0)}/z'$ is real on the boundary, where $'$ denotes
the 
derivative with respect to some parameter along the boundary, and by using
$A=g(h^{-1}\p_zh)g^{-1}z' -g'g^{-1}$, as follows
\be\label{liftdiff}
\frac{\CL_{\xi}\sigma}{\sigma}= -\frac{1}{4\pi i}\oint\tr(g^{-1}(\xi
g)g^{-1}\rd g ) + \frac{1}{4\pi
i}\int\tr[(h^{-1}h_{z})^{2}]\db\xi^{(1,0)}\wedge\rd z.
\end{equation}
where we have put $\CL_\xi \sigma = \nabla_{X}\sigma -i
H_{\xi}(A)\sigma$.
\begin{remark}
This action is not holomorphic in general since the integral over
    $D_{+}$ does not depend holomorphically on $A$ (indeed the action
    on $\CA_{+}$ was not holomorphic either). The prequantum operator
    therefore does not send polarised (holomorphic) sections to
    polarised sections and more work needs to be done to quantize the
    action of a general diffeomorphism.
\end{remark}
We will not, however, be interested in the action of general vector
fields on the disc, but of $\hV(D_+)$
(i.e.\ real vector
fields $\xi$ for which $\xi^{(1,0)}$ is holomorphic).  We have
\begin{proposition}  
There exists a holomorphic action of $\hV(D_{+})$ on the determinant
line bundle given by
\begin{equation}\label{basic3}
\frac{\CL_{\xi}\sigma}{\sigma}= -\frac{1}{4\pi i}\oint\tr(g^{-1}(\xi
g)g^{-1}\rd g) 
\end{equation}
\end{proposition}
\begin{proof}
  This is not completely trivial as vector fields in $\hV(D_{+})$ are
  not in general tangent to the boundary.  We get around this by
  representing $\hV(D_{+})$ as a quotient of complexified vector
  fields that are tangent to the boundary whose $(1,0)$-part is
  holomorphic, by complexified vector fields, tangent to the boundary,
  whose $(1,0)$-part is zero.  The formula can be extended to complex
  vector fields (i.e.\ $\xi=\xi^{(1,0)}\p_z+\xi^{(0,1)}\p_{\bz}$ with
  $\xi^{(1,0)}$ independent of $\xi^{(0,1)}$, except on the boundary
  where $\xi^{(1,0)}\bz'=\xi^{(0,1)}z'$) by requiring complex
  linearity.  We can now restrict to the Lie algebra of complex vector
  fields such that $\xi^{(1,0)}$ is holomorphic on $D_+$.  This latter
  subalgebra acts holomorphically as it is defined by the condition
  $\db\xi^{(1,0)}=0$, and so our formula simplifies to the boundary
  integral \eqref{basic3} analogous to the case for complex gauge
  transformations.  Thus, naive geometric quantization does quantize
  the action of this algebra of complex vector fields tangent to $\p
  D_+$ with holomorphic $(1,0)$ part.
  
  To obtain an action of $\hV(D_+)$, observe that the subalgebra of
  complex vector fields of the form $\xi^{(0,1)}\p_{\bz}$, with
  $\xi^{(0,1)}|_{\p D_+}=0$, is a Lie algebra ideal, acts trivially and
  the quotient of the algebra of complex vector fields tangent to $\p
  D_+$ and with $\xi^{(1,0)}$ holomorphic by this subalgebra is
  $\hV(D_{+})$.  We have therefore obtained the desired holomorphic
  action of $\hV(D_{+})$ on the determinant line bundle.
\end{proof}

\setcounter{equation}{0}
\section{Reduction of phase space}\label{red}
All the constructions of the last two sections are invariant under the
action of $\CG_0$ and the final formulae are invariant under the
action of $\CG^c_0$; it is therefore natural to attempt to reformulate
the constructions in terms of a reduced phase-space obtained by the
symplectic reduction in the case of $\CG_0$ or the straight quotient
in the case of $\CG_0^c$.  In the following we consider the case when
$D_+$ is just a disc.  Then these reductions both give the same
reduced phase space $\CAr=LSU_n/SU_n$, where
$LSU_n=$map$\{S^1,SU_n\}$ is the loop group of $SU_n$, as a
consequence of one of the factorization theorems for loop groups. When
$D_+$ is the complement of more than 1 disc, the symplectic and
complex quotients are still the same, and the general theory follows
in much the same way, but the connection with loop groups is less
relevant.  The connection with loop groups arises as follows (see
Donaldson 1992 for a full discussion).

From the complex point of view, the reduced phase-space
$\CAr=\CA_+/\CG^c_0$. This follows by
noting that if $\db_\alpha\in \CA_+$, then there exists a complex gauge 
transformation $F$, smooth up to the boundary, such that
\begin{equation}\label{cpxg}
\db_\alpha = F\cdot \db\cdot F^{-1}.
\end{equation}
This $F$ is not unique; the group $L^+SL_n(\C)$ of
holomorphic maps $D_+ \to SL_n(\C)$ that are smooth up to the boundary
acts by multiplication on the right, $F\mapsto FG $ preserving
\eqref{cpxg}. Elements of $\CG_0^c$ act on the left leaving invariant
the boundary value of $F$.  Furthermore, any two $F$'s with the
same boundary values are related by an element of $\CG_0^c$. Hence
$\CA_+/\CG^c_0$ is the homogeneous space $L SL_n(\C)/L^+ SL_n(\C)$.

The symplectic reduction of $\CA_+$  is obtained 
by dividing the zero-set of the moment map $\mu:\CA_+\mapsto T^*_e\CG_0$
$$
\langle u, \mu(A)\rangle = H_u(A)
$$
for the action of $\CG_0$ on $\CA_+$, by $\CG_0$. Note that the
boundary term in $H_u(A)$ is absent if $u\in\mfg_0$, so the zero-set
of $\mu$ consists exactly of the flat unitary connections on $D_+$.
Any such connection is gauge equivalent to the trivial connection $\rd$,
\begin{equation}
    \rd_A = \gamma\cdot \rd\cdot \gamma^{-1}.
\end{equation}    
This time, $\gamma\in \CG$ is 
determined up to the action of the
gauge transformations, $\gamma \mapsto \gamma\eta$, where 
$\eta$
is constant. The group $\CG_0$ acts on the left, fixing the boundary 
value of $\gamma$. Hence $\mu^{-1}(0)/\CG_0$ is identified with 
$LSU_n/SU_n = \Om  SU_n$.

In the theory of loop groups, Pressley \& Segal (1986), we have that 
$$
LSL_n(C)/L^+SL_n(\C) = L SU_n/SU_n\, ,
$$
i.e., for any $F$ in $LSL_n(\C)$ there is a $G$ in $L^+ SL_n(\C)$
unique up to constants such that $FG$ is unitary.  This shows that the
real and complex reductions give the same answer when $D_+$ is
a standard disc.

The equivalence between the symplectic and complex reduction is true
also in our more general situation when the boundary of $D_+$ is a
disjoint union of circles.

\begin{remark} If we consider the smaller phase space of 
$\db$-operators with compact support in $D_+$, then this 
correspondence fails. Indeed a flat connection with compact support 
is gauge equivalent to $\rd$ by a gauge transformation that is 
necessarily constant near $\p D_+$; so all such flat connections are 
in one orbit of $\CG_0$.
\end{remark}

We shall now reduce all the objects so far considered by $\CG_0$. For
each object (e.g.\ symplectic form, Hamiltonian, prequantum data) we
shall restrict to the zero-set of $\mu$ and then check that this
restriction is $\CG_0$-invariant. The end result are explicit formulae
for these reduced objects in terms of the loop space $\Om SU_n$.

We first give more details of the identification between 
$\CA^{\rm red}$ and $\Om SU_n$. 
Fix $A\in\mu^{-1}(0)$, so that $A$ is a flat connection in
$\CA_+$. Then there exists $\gamma \in \CG$, unique up to
multiplication by a (constant) element of $SU_n$, with $\rd_A = \gamma\cdot
\rd \gamma^{-1}$ so that
$$
\mu^{-1}(0) = \CG/SU_n
$$
where $SU_n$ here denotes the subgroup of constant gauge
transformations.  Thus any flat infinitesimal deformation of $A$ is given
by a gauge transformation and so 
\begin{equation} \label{tgt23}
T_A\mu^{-1}(0) = \{-\rd_A u: u\in \mfg=T_e\CG\}.
\end{equation}
Note that the tangent space of $\CG$ at a point $\gamma$ is
identifiable with the Lie algebra $\mfg$, $\mfg \mapsto T_\gamma\CG$,
by $u \mapsto u\gamma$ and this is compatible with \eqref{tgt23} (in
physicists' notation $\delta_u\gamma\,\gamma^{-1} =u$). One can check
that the left-action of $\CG$ on $\CG/SU_n$ coincides with the gauge
action of $\CG$ on $\mu^{-1}(0)$.

Inside $T_A\mu^{-1}(0)$, there is the vertical tangent space $T_A^V$
which is generated by the $\CG_0$-gauge orbits. By definition
$$
T_A ^V =\{-\rd_A u: u\in \mfg_0\}.
$$
The quotient $T_A\mu^{-1}(0)/T_A ^V=T_{[A]}\CAr$,
is thus identified with the space of boundary values $-\rd_A
u|{\p D_+}$.  The exact sequence of tangent spaces
$$
0 \to T_A ^V \to T_A\mu^{-1}(0) \to T_{[A]}\CAr \to 0
$$
gets identified with the exact sequence
$$
0 \to \mfg_0 \to \mfg/{\mathfrak su}_n \to L{\mathfrak
su}_n/{\mathfrak su}_n \to 0.
$$

We now compute the reduced symplectic form
$\Omr$ on $\CAr$. With the above identifications, we take tangent 
vectors $-\rd_A u$ and $-\rd_A v$ at $A$ corresponding to elements 
$u,v\in \mfg$. Then
\begin{equation}\label{bot}
\Om(d_\gamma u,d_\gamma v)= \frac{1}{2\pi}\int\tr(d_\gamma u\wedge
d_\gamma v) = \frac{1}{2\pi}\oint \tr(u\,d_\gamma v) = -\frac{1}{2\pi}
\oint \tr(v\,d_\gamma u)
\end{equation}
where here and below we have used $u$ and $v$ to denote $u|_{\p D_+}$
and $v|_{\p D_+}$ respectively.  This is equivalent to the standard
symplectic form on $\CAr$ from Pressley--Segal (1986) (p. 147). There
this form is written down at the identity coset in $\CAr$ and
propagated as a left-invariant form over the whole of $\CAr$.  This
agrees with the above formula since, starting with the Pressley--Segal
definition,
$$
\Omr_\gamma(u\gamma,v\gamma) = \Omr_1(\gamma^{-1}u\gamma,
\,\gamma^{-1}v\gamma) = 
\int\tr(\gamma^{-1}u\gamma\,\rd(\gamma^{-1}v\gamma)) = 
\int\tr(u\rd_\gamma v)$$
and the last expression is exactly as in \eqref{bot}.

Similarly for the Hamiltonians we find, for the action of gauge
transformations, 
$$
\Hr_u(\gamma) = -\frac{1}{2\pi}
\oint\tr(d\gamma\, \gamma^{-1}u)
$$
from \eqref{hamf1} and for the action of diffeomorphisms from \eqref{hamf2}
$$
\Hr_\xi(\gamma) = -\frac{1}{4\pi}\oint 
\tr(\rd\gamma\gamma^{-1}\,\xi(\gamma)\gamma^{-1}).
$$
To summarize,
\begin{proposition}
The reduced symplectic form $\Omr$ on $\CAr$ is given by
$$
\Omr_\gamma(u,v) = \frac{1}{2\pi}\oint\tr(u\,d_\gamma v).
$$
The action of $LSU_n$ upon $\CAr$ is generated by the Hamiltonian
$$
H_u(\gamma) = - \frac{1}{2\pi}\oint\tr(d\gamma\,\gamma^{-1}u);
$$
the action of $\rV(S^1)$ is generated by the Hamiltonian
$$
H_\xi(\gamma) = \frac{1}{4\pi}\oint\tr(d\gamma\,\gamma^{-1}\cdot
\xi(\gamma)\gamma^{-1}).
$$
\end{proposition}
\begin{remark} The phase-space $\CAr$ is that for
  Wess-Zumino-Witten theory in two dimensions.  The Lagrangian does
  not have a local, invariantly defined formula, but its first
  variation does:
$$
\delta L=\int_D \tr ( g^{-1}\delta g  \rd (g^{-1} \p g)).
$$
The symplectic form and Hamiltonians can be derived from this
Lagrangian by standard applications of Noether's theorem.
\end{remark}

\subsection{Quantization of the reduced space}

We now proceed to analyze the reduction of the prequantum data, i.e.,
the Quillen connection on $\Det$ and the lifts of actions of gauge
transformations and diffeomorphisms for the reduced phase space. 

We first simplify Quillen's formula for $J_A$ in
\eqref{derivfrm}. Writing $\rd_A =\gamma\cdot\rd\cdot\gamma^{-1}$ for
the flat connection $A$ as before, we have
$\alpha^*=-\p\gamma\,\gamma^{-1}$ but to determine $\beta$ we need the
gauge 
transformation $g$ in \eqref{gg}. This amounts to solving
$$
\gamma\cdot\db\cdot\gamma^{-1} = g\cdot\db\cdot g^{-1}\mbox{ in }D_+,\;
\db = g\cdot\db\cdot g^{-1}\mbox{ in }D_-.
$$
(Strictly, we should have $\db_0$ here.) From these equations, $g$ is 
holomorphic in $D_-$ and $g^{-1}\gamma$ is holomorphic in $D_+$. Thus 
$g$ is determined by the Birkhoff factorization 
\begin{equation}\label{birk}
\gamma = g\,g_+^{-1} \quad \mbox{ on $\p D_+$,}
\end{equation}
 where $g_+\in L^+ SL_n(\C)$, $g\in L^- SL_n(\C)$; $g_+$
and $g$ are the positive and negative-frequency parts of $\gamma$ and
$g$ is then continued over $D_+$ by requiring \eqref{birk} to hold on
$D_+$. 

Using this we obtain
$$
\beta-\alpha^* = \p g\, g^{-1} + \p \gamma \,\gamma^{-1}
$$
or in terms of $g_+$ and $\gamma$, 
$$
\beta - \alpha^* = \gamma(\p g_+\, g_+^{-1})\gamma^{-1}.
$$
Now if $u\in \mfg$ is used as before to define a tangent vector 
$-\rd_A u$ to $\mu^{-1}(0)$, we get the formula
$$
\frac{\nabla_u\sigma}{\sigma}= -\frac{1}{2\pi}\int\tr[
\gamma(\p g_+\, g_+^{-1})\gamma^{-1} \wedge(-\rd_A u)].
$$
To simplify this, note that we can replace $\rd_A$ by $\db_A$; 
writing this derivative out in terms of $\gamma$, we get
$$
\frac{\nabla_u\sigma}{\sigma}= \frac{1}{2\pi}\int\tr
(\p g_+\, g_+^{-1}) \wedge(\db(\gamma^{-1}u\gamma)).
$$
Since $g_+$ is holomorphic in $D_+$ the obvious integration by parts 
reduces this to a boundary integral,
\begin{equation} \label{desc}
\frac{\nabla_u\sigma}{\sigma}= -\frac{1}{2\pi}\oint\tr(
(\rd g_+\, g_+^{-1}) \gamma^{-1}u\gamma) =
-\frac{1}{2\pi i}\oint\tr[
(\rd g\,g^{-1} - \rd\gamma\,\gamma^{-1})u].
\end{equation}
This vanishes if $u\in \mfg_0$; it follows that the prequantum data 
descends to $\CAr$. Moreover, the vacuum-state $\sigma$ also descends
to define a vacuum state in the reduced theory. We shall 
denote this by $\sigma$, rather than by $\sigma^{\rm
red}$. Notice that \eqref{desc} depends only on the boundary
value of $u$ and $\gamma$, so it could be used as a definition of
pre-quantum data over $\CAr= \Om SU_n$. We have proved

\begin{proposition} The symplectic reduction of Quillen's connection 
yields a connection on the determinant line-bundle over $\Om SU_n$ 
whose curvature is $\Omr$. 
\end{proposition}
Although this is probably well known ($\CAr$ and its
determinant bundle have been well studied) we are not aware of a
previous occurrence of such a result in the literature. It gives a
direct link between Quillen's construction and the determinant
line-bundle over $\Om SU_n$. The latter is usually defined by
expressing $\Om SU_n$ as a grassmanian and restricting the determinant
line-bundle of the grassmanian of Hilbert space.

Using the same recipe as before for the lift of an action, we find 
that the variation of $\sigma$(=$\sr$) under the action of 
$LSU_n$ on $\Om SU_n$ is given by:
\begin{equation}\label{basic4}
\frac{\CL_u\sigma }{\sigma}
= -\frac{1}{2\pi i}\oint (\rd g g^{-1} u)
\end{equation}
where $g$ is the `negative-frequency part' of $\gamma$. Similarly for 
diffeomorphisms, we get 
$$
\frac{\CL_\xi\sigma}{\sigma} =
-\frac{1}{2\pi i}\oint \tr[\rd g g^{-1} \xi\gamma\gamma^{-1}
- \half \rd\gamma\,\gamma^{-1}\xi\gamma\,\gamma^{-1}].
$$
This can be simplified almost exactly as for the previous
calculation for the action of the diffeomorphism group:
cf. \eqref{horr1} and the ensuing calculations. One obtains, finally,
\begin{equation}\label{basic2}
\frac{\CL_\xi\sigma}{\sigma} =-\frac{1}{4\pi i}\oint\xi\,d\theta\tr[(g'g^{-1})^2 - (g_+'g_+^{-1})^2].
\end{equation}
As before, this does not preserve the complex structure in general 
(for $g_+$ does not depend holomorphically on $\gamma$) but when $\xi$ 
is the boundary value of a holomorphic vector field in $D_+$, 
the latter term drops out, leaving a holomorphic lift.

\section{Construction of the Fock space}\label{fock}
The space $\CH_1$ of all holomorphic sections of $\Det$ over $\CA_+$
will be too large for the Fock space as we only wish to consider
sections that are invariant under $\CG_0$.  The space $\CH_0$ of
holomorphic sections of $\Det$ over $\CAr$ is the space of such
invariant sections, but will now be too large because we need
distinguish the subspace $\CH$ of `square-integrable' holomorphic
sections.  We shall obtain $\CH \subset \CH_0$ as the completion of a
dense subset $\CC\subset \CH_0$.  The set $\CC$ will be a set of
coherent states (which, in particular, is not a linear subspace of
$\CH_0$).  The inner product is constructed using Segal's gluing
formula for determinants. The following discussion of these
matters runs roughly parallel to that of \S10 of Segal and Wilson
1985.

Let $\CP^1 = D_+ \cup D_-$ as before, and assume that $D_{\pm}$ are
standard discs with common boundary $S^1=\{z\in\C ,\, |z|=1\}$. Let
$\rho$ be the anti-holomorphic involution of $\CP^1$ given by
reflection in $S^1$, $z\rightarrow 1/\bar{z}$. Since $\rho$ switches
$D_+$ and $D_-$, the assignment $\overline{\alpha} = \rho^*(\alpha^*)$
maps endomorphism-valued $(0,1)$-forms on $D_+$ to endomorphism-valued
$(0,1)$-forms on $D_-$. This gives an anti-holomorphic map $\db_\alpha
\mapsto \db_{\bar{\alpha}}$ from the space $\CA_+$ to the space
$\CA_-$ of $\db$-operators over $D_-$.

Given now $\db$-operators $\db_\alpha$, $\db_{\alpha'}$, we denote by
$\db_{\overline{\alpha}\cup \alpha'}$ the $\db$-operator over $\CP^1$
which is equal to $\db_{\overline{\alpha}}$ in $D_-$ and to
$\db_{\alpha'}$ over $D_+$.  In general this has a jump-discontinuity
along $S^1$ (cf.\ \S\ref{ujmp}). It can be shown (Segal 1989 \& 1991)
that the determinant lines of the $\db$-operators are related by
$$
\Det(\db_{\overline{\alpha}}) = \overline{\Det(\db_\alpha)},\;\;
\Det(\db_{\overline{\alpha}\cup\alpha'}) = 
\overline{\Det(\db_\alpha)}\otimes \Det(\db_{\alpha'}).
$$
In particular 
$$
\Det(\db_{\overline{\alpha}\cup \alpha})=
 \overline{\Det(\db_\alpha)}\otimes\Det(\db_\alpha)
$$
is canonically the complexification of an oriented real line; the real
positive elements are those of the form $\overline{u}\otimes u$, for 
$0\not= u \in \Det(\db_\alpha)$. Furthermore, we have
$$
\det(\db_{\overline{\alpha}\cup \alpha }) > 0
$$
for every $\alpha$.

For fixed $\alpha$, 
$$
\det(\db_{\overline{\alpha}\cup \alpha'}) \in 
\overline{\Det(\db_\alpha)}\otimes\Det(\db_{\alpha'})
$$
depends holomorphically on $\alpha'$ and so defines a ray in
$\CH_0$. A genuine state arises by fixing a non-zero element
$\Psi(\alpha)\in\Det^*(\db_\alpha)$; since it depends
anti-holomorphically on $\alpha$, it is natural to write it as a `bra'
$\langle \Psi(\alpha) |$. The operation of evaluation at
$\alpha'$ gives an element of $\Det(\alpha')$ so that given
$\Psi(\alpha')\in \Det^*(\alpha')$ we can obtain a complex number.
This operation can be thought of as evaluation of the `bra' against
the `ket' $|\Psi(\alpha')\rangle$, and we have the suggestive
formula
$$
\langle \Psi(\alpha) |\Psi(\alpha')\rangle =
\det(\db_{\overline{\alpha}\cup\alpha'})\overline{\Psi(\alpha)}\Psi(\alpha')\in
\C \, , 
$$
and $\langle \Psi(\alpha) |\Psi(\alpha)\rangle$ is positive
definite.  We see that $\Psi(\alpha)$ plays the role of a type of
coherent state corresponding to $\db_\alpha$.

We can now define $\CF$ to be the completion inside $\CF_0$ of the
linear span of all the $\Psi(\alpha)$'s with respect to the inner
product $\langle\Psi(\alpha)|\Psi(\alpha')\rangle $.
This is the Fock space for our theory.  Note that $\CF$
will not be dense in $\CF_0$ as all elements of $\CF$ are
invariant under the group of based gauge transformations, and this
will certainly not be the case for $\CF_0$. 

Finally we can give the quantum-field-theoretic interpretation of 
formula \eqref{deftau1} for the $\tau$-function. The holomorphic section
$\sigma$ corresponds to the vacuum `bra' $\langle \Psi(0)|$, the
coherent state based on $\db_0$.  A family of elements $G(t)$ of
either the gauge or diffeomorphism group depending on a parameter $t$
acts on a coherent state vector by sending $|\Psi(\alpha)\rangle$ to
$|\widehat{G(t)}\Psi(\alpha)\rangle$ which is the coherent state based
on $G(t)\alpha$ with $\widehat{G(t)}\Psi(\alpha)\in
\Det^*(\db_{G(t)\alpha})$ given by the lifted action of $G(t)$ on
$\Det$.  In this context, then,
$$
\tau= 
\langle \Psi(0)|
\widehat{G(t)}\Psi(\alpha)\rangle=
(\sigma(G(t)\alpha),  \widehat{G(t)}\Psi(\alpha) ) 
=(\widehat{G(t)}{} ^{-1}\sigma(G(t)\alpha),  \Psi(\alpha) ) 
$$
where the last two pairings are between elements of $\Det$ and
$\Det^*$.  Modulo the irrelevant extra constant factor of
$\Psi(\alpha)$, this is \eqref{deftau0} and differentiation with
respect to the parameter $t$ leads to the definition \eqref{deftau1}.
If we now consider the case where the submanifold of $\CG^c$ is an
abelian subgroup with Lie algebra generators $\phi_1$ and $\phi_2$,
then with parameters $(x,t)$ we have $g(x,t)=\exp (x\phi_1+t\phi_2)$ and
we have arrived at the QFT formula \eqref{japfrm}.


\setcounter{equation}{0}
\section{Integrable equations and their twistor description}
\label{s2}
In this section we review the twistor correspondence for the Bogomolny
equations and its reductions appropriate to the KdV equation and the
Ernst equations.  Although a wider variety of integrable systems can
be obtained by considering reductions of the self-dual Yang-Mills
equations, reduction of the Bogomolny equations yields many of the most
famous examples, the KdV equations, the Ernst equations, the Sine
Gordon equation, the nonlinear Schrodinger equation, sigma models and
so on.  The aim is to apply the technology of \S\ref{gq} and
\S\ref{red} to obtain the $\tau$ function in terms of the twistor data
in the next section.  To do this, our goal in this section is
\S\ref{lift} where we reformulate the twistor construction so that it
transforms a solution to an integrable system into a bundle with $\db$
operator (or patching function) on a family of $\CP^1$s parametrized
by space-time such that the $\db$ operator or patching function
changes by a combination of a gauge transformation and a
diffeomorphism as the point in space-time changes.  To obtain such a
description in a completely natural way, we will need to impose a
symmetry on the Bogomolny equations.  This will allow us to define the
tau function to be the determinant of the $\db$ operator on each of
the $\CP^1$'s in the next section.

Some of this material duplicates that in Mason, Singer \& Woodhouse
(2000) and Mason \& Woodhouse (1996) but is included to make the
present paper more self-contained. For readers of Mason, Singer \&
Woodhouse (2000), we point out that Proposition 6 and the key equation
(14) of that paper corresponds to our formulae \eqref{deftau1} using
either the pair of formulae
\eqref{basic1} \& \eqref{basic3} or  \eqref{basic4} \& \eqref{basic2}.

\subsection{The geometry of the twistor correspondence}
We will here work just with the correspondence between $\C^3$ and its
two-dimensional minitwistor space.  This framework has several
important generalizations, see, for example, Mason et.\ al.\ (2000) for
details. 

Space-time $\CM$ will be taken to be 
$\C^3$ with coordinates $(t,x,v)$ and  metric $dx^2 -
2dv\odot dt$.  Twistor space $\CZ$ is defined to be the space of 
complex null 2-planes in $\C^3$ (i.e.\ the restriction of
the metric to the 2-plane should be degenerate). Every such null plane
is spanned by a pair of vector fields of the form
$$
V_0= \p_x - \lambda \p_v \, , \quad V_1= \p_t - \lambda \p_x
$$
for some $\lambda$ (possibly equal to $\infty$) and is orthogonal
to the null direction $V_1-\lambda V_0$. The twistor space $\CZ$ can
be represented as the total space of the complex line bundle $\CO(2)$
of Chern class 2 over $\CP^1$ (which is the tangent bundle $T\CP^1$).
Using a fibre coordinate $\mu$, and affine 
coordinate
$\lambda$ on $\CP^1$, the correspondence with space-time is
\begin{equation}\label{tcorr}
\mu = v + \lambda x + \lambda^2 t\, ;
\end{equation}
it is easily checked that, for fixed $\mu$ and $\lambda$, this
equation determines a null 2-plane in $\C^3$ and that all null two
planes arise in this way if we introduce coordinates $(\tl,\tm)$,
where $\tl = \lambda^{-1}, \quad \tm = \lambda^{-2}\mu$ to cover the a
neighbourhood of $\lambda=\infty$.  [The fibre of $\CO(2)$ over
$\lambda =\infty$ is distinguished in the KdV example and the
coordinates $(\tl,\tm)$ will be used for calculations there.]

Alternatively, if we fix a point $p\in\CM$ with coordinates $(v,x,t)$
in \eqref{tcorr}, we obtain a rational curve in $\CZ$, a holomorphic
cross-section of $\CO(2)$, which we denote by $L_{(v,x,t)}$ or $L_p$.
Thus we have a correspondence wherein the points of $\CZ$ parameterize
the null 2-planes in $\C^3$, and the points of $\C^3$ parameterize the
cross-sections of $\CZ$. These cross-sections will sometimes,
incorrectly, be referred to as twistor lines (being, in fact,
conics).

The correspondence can be summarized by the double fibration of the
correspondence space $\CF=\C^3\times \CP^1=\{(p,Z)\in\CM\times\CZ |
Z\in L_p\}$ (which, more invariantly is the projective spin bundle or
bundle of null directions over $\CM$) over space-time and twistor
space:
$$\begin{array}{rccclcrcccl}
&&\C^3\times\CP^1&&&&&&\CF&&\\
&p\swarrow&&\searrow q&&=&&p\swarrow&&\searrow q&\\
\C^3&&&&T\CP^1&& \CM &&&&\CZ
\end{array}
$$
The fibres of the projection $q$ are spanned by the vector
fields $V_0$ and $V_1$, and the projection $p$ is the projection onto
the first factor.

\subsection{The Bogomolny equations and the Ward correspondence}
\label{wcx}
For our purposes, the Bogomolny equations are best defined to be the
integrability condition for the Lax pair
\begin{equation} \label{lp1}
L_0= (\p_x + A) - \lambda (\p_v + B )\, , \quad L_1= (\p_t + C) -
\lambda (\p_x +D) 
\end{equation}
where the independent variables $(v,x,t)$ are coordinates on $\C^3$,
and the dependent variables $A, B, C$ and $D$ are functions on $\C^3$
with values in the Lie algebra of some gauge group, which will be
$SL(2,\C)$ in the examples we will consider. In the context of Lax
pairs, the affine Riemann sphere coordinate $\lambda$ is more commonly
known as the `spectral parameter'.  More invariantly, one should think
of $L_0$ and $L_1$ as differential operators on a trivial bundle over
the correspondence space $\CF$.  We will be interested in
gauge-equivalence classes of such operators.

The natural symmetry group of the equations is the complex Euclidean
group together with dilations associated to the metric $\d s^2=\d x^2
- 2\d v\odot \d t$.  Later we will see that symmetries can be imposed
so that the equations reduce to the KdV equation or the Ernst
equation.



The Ward correspondence provides a 1:1 correspondence between
solutions to the
$SL(2,\C)$ Bogomolny equations on $\C^3$, and rank-2
holomorphic vector bundles $E \to \CZ$ such that $E$ is trivial over
each twistor line.\footnote{More generally we can restrict the domain
  to some Stein open set $U\subset\C^3$ and have such a correspondence
  with bundles over $q(p^{-1}(U)\subset\CZ$ trivial over the $\CP^1$'s
  corresponding to points of $U$.}
This result is standard and will not be proved here (see for example
Ward \& Wells 1990, Mason \& Woodhouse 1996). We shall, however,
need some details of the correspondence and so recall briefly how it
works.   

\medskip
\noindent 
{\bf Obtaining a bundle on twistor space from a solution to the
  Bogomolny equations:} Given a solution to the Bogomolny equations on
$\C^3$, we introduce the associated Lax pair $L_0$ and $L_1$ as in
equation (\ref{lp1}).  We define a fibre $E_Z$ for $Z\in\CZ$ of the
holomorphic vector bundle $E\to\CZ$ to be the space of solutions to
the Lax pair over the null-plane in $\C^3$ corresponding to $Z$.  (The
integrability conditions ensure that $E_Z$ is a complex 2-dimensional
vector space.) 
To be consistent with our
subsequent conventions, the matrices in the Lax pair will be assumed
to be acting on the right.

\medskip
\noindent
{\bf Obtaining a solution to the Bogomolny equations from a bundle on
twistor space:} 
Suppose we are given a holomorphic bundle $E\to \CZ$, trivial
on each twistor line. Pull $E$ back to $\CF$, to obtain a bundle
$\hE=q^*E$ that is canonically trivial over the fibres of $\CF \to
\CZ$. The pair $V_0$ and $V_1$ of vector fields are tangent to these
fibres and the canonical triviality means that these have canonical global holomorphic
lifts to the bundle, $L_0$ and $L_1$.  We  have assumed that $E$ is
trivial over each $\CP^1$ in $\CZ$ so that $\hE$ is trivial over each
$\CP^1$ fibre of $\CF=\C^3\times\CP^1$ over $\C^3$ and so we can
trivialize $\hE$ over $\CF$.  In such a
trivialization $L_0$ and $L_1$ will be holomorphic in $\lambda$ 
with a simple pole at $\lambda=\infty$ and so must take the form as
given in (\ref{lp1}).  Since $(L_0,L_1)$ are gauge equivalent to
$(V_0,V_1)$ in a frame pulled back from $\CZ$, they must commute.

To make this transform more explicit, we must first choose one of the 
following explicit presentations of $E \to \CZ$.

\noindent
{\bf Cech presentation:} Cover $\CZ$ with two open sets,
$U_\pm=\{(\mu,\lambda)| \; |\lambda|^{\pm 1} < 1+\varepsilon \}$ so 
that
$E$ is trivial over $U_+$ and $U_-$. The bundle is then completely 
described by the transition function (patching function) 
defined on
$U^+\cap U^-$.
On $\CF$, the pull-back $\hP$ of the patching matrix
$P$ defining $E$ is annihilated by $V_0$ and $V_1$. On the other hand,
the assumed holomorphic triviality over each twistor line means that
$\hE$ is trivial on $\CF$ so there exist maps $g_\pm(\lambda;v,x,t)$
holomorphic on $\pi_1^*U_\pm$ respectively with $\hP = g_-g_+^{-1}$
defining a global frame of $\hE$ over $\CF$.  Operating with $V_0$ and
$V_1$ gives, using $V_iP=0$,
\begin{equation}\label{lpd}
g_-^{-1} V_ig_- = g_+^{-1}V_ig_+=L_i
\end{equation}
so that the $L_i$ are global over each Riemann sphere. From the form of
$V_0$ and $V_1$, the  $L_i$ are holomorphic in $\lambda$ with a simple pole at
$\lambda=\infty$; hence they are linear in $\lambda$ and have the form of a
Bogomolny Lax pair (\ref{lp1})
 thus defining $A, B, C, D$ as functions
only of $(v,x,t)$ as in (\ref{lp1}).  Moreover, \eqref{lpd} implies
that $L_ig_\pm=0$ which in turn implies that the $L_i$ commute so that
$A, B, C, D$ satisfy the Bogomolny equations.

\noindent
{\bf Dolbeault presentation:} 
Topologically $E$ is a trivial bundle, so we can regard it as
the product bundle $\C^2\times \CZ$, equipped with a non-trivial
$\db$-operator $\db_\alpha = \db + \alpha$, where $\alpha$ is a
$(0,1)$-form with values in ${\mathfrak sl}(2,\C)$, as in
\S\ref{condb}, satisfying the integrability condition
$\db_\alpha^2 =0$; in full,
$$
\db\alpha + \alpha\wedge\alpha = 0.
$$
In this description of holomorphic bundles, there is much freedom in
the choice of smooth identification of $E$ with $\C^2$. This
translates into a gauge freedom in $\alpha$, so that $\db_\alpha$ and
$\db_{g(\alpha)}:= g\db_\alpha g^{-1}$ define equivalent holomorphic
bundles. Explicitly the complex gauge transformations $g: \CZ \to
SL(2,\C)$ act on the space of $\db$-operators by the formula
$$
g(\alpha) = g\alpha g^{-1} - \db g\, g^{-1}.
$$
In the case of holomorphic bundles over $\CZ$, each fibre of
$\CZ\to\CP^1$ is
Stein so that one can choose a gauge which is
holomorphic in the fibre direction and so $\alpha$ can be reduced to 
$a\,d\bl$. The integrability condition immediately shows that $a$ is
holomorphic in $\mu$. This will be a convenient `partial gauge fixing'
in what follows; $g$ must now be holomorphic in $\mu$.

We now repeat the previous operations using $\db_\alpha$ instead of
the patching description.  we pull back the $\db$-operator to $\CF$,
and denote the pullback by the same symbol $\db_\alpha$.  This
operates on the product bundle $\C^2\times \CF$ pulled back from
$\CZ$. This operator commutes with the action of $V_0$ and $V_1$ on
$\C^2\times \CF$; and the assumption of holomorphic triviality on each
line implies that there also exists a gauge transformation $g:\CF \to
SL(2,\C)$ such that $g\db_\alpha g^{-1}= \db$, the trivial
$\db$-operator on $\C^2\times \CF$.  Now define
$$
L_0= g V_0g^{-1}, L_1 = g V_1 g^{-1}.
$$
These commute with $g\db_\alpha g^{-1}= \db$, so are holomorphic on
$\CF$. On the other hand $L_0$ and $L_1$ have simple poles at $\infty$
in $\lambda$ and so are linear in $\lambda$. Hence they are in the form of
the Lax pair (\ref{lp1}) and clearly commute with each other as
$g^{-1}$ is a solution (on the right).

\subsection{Imposition of symmetries}
\label{impsymm}
The naturality of the Ward transform ensures that if a solution to the
Bogomolny equations admits a symmetry, then the corresponding
holomorphic vector bundle on twistor space is invariant under a
corresponding motion of $\CZ$.  Such symmetries must be taken from the
group $G$ consisting of complex translations and the conformal
orthogonal group $CO(3,\C)$. Any element of $G$ permutes the null
2-planes so that $G$ acts naturally on $\CZ$ and $\CF$ by biholomorphic
transformations. Thus any generator $X$
of $G$ corresponds to holomorphic vector fields $\tX$ on $\CZ$ and
$\hX$ on $\CF$.


The most general holomorphic vector field defined globally on $\CZ$ is
given by
$$
\tX = (\alpha \lambda^2 +\beta\lambda +\gamma)\p/\p\lambda + ((2\alpha
\lambda + \beta + \delta) \mu +
a\lambda^2 +b\lambda +c)\p/\p \mu.
$$
and the reader may verify that this corresponds to 
$$
X =  a\p_t + b\p_x + c\p_v +
     \alpha(2v\p_x+ x\p_t) + \beta(v\p_v -t\p_t) -
     \gamma( x\p_v+2t\p_x) + \delta(v\p_v + x\p_x +t\p_t)
$$
on $\C^3$ and to
$$
\hX =  X + V
$$
on $\CF$, where
$$
V = (\alpha\lambda^2+ \beta\lambda + \gamma)\p_\lambda
$$
on $\CF$. We see from this that the parameters $a$, $b$, $c$ yield
translations of $\C^3$, $\alpha$, $\beta$, $\gamma$ yield (complex)
rotations while $\delta$ yields the dilation.

Let $X$ be the generator of an infinitesimal symmetry of $\C^3$ as
above. A Lax pair (\ref{lp1}) will be $X$-invariant if there exists
Lie derivative operator $\CL_\hX$, which commutes with the Lax pair as
follows:
$$
[\CL_\hX, L_0] = \alpha \lambda  L_0  -\alpha L_1\, , \qquad
[\CL_\hX, L_1] = \gamma  L_0  + (\alpha \lambda +\beta)L_1\, . 
$$
Here $\CL_\hX$ is by definition a linear differential operator on the
sections of the bundle over $\CF$ with the property
$$
\CL_\hX(f\otimes s) = (\hX f)\otimes s + f\otimes \CL_\hX s
$$
whenever $f$ is a function and $s$ is a section. It follows that in
any local gauge, $\CL_X$ takes the form
$$
\CL_\hX = \hX + \Phi
$$
for some matrix function $\Phi$. We also assume in this discussion
that $\Phi$ is holomorphic in any holomorphic gauge.

In terms of $\CL_\hX$, the local invariant gauge is one in which
$\CL_\hX = \hX$, and then the commutation conditions show that $XA =0$
etc.  It is not difficult to show that the Ward transform restricts to
give a 1:1 correspondence between $\tX$-invariant bundles over $\CZ$
and $X$-invariant solutions of the Bogomolny equations, whenever $X$
and $\tX$ are corresponding vector fields on $\C^3$ and $\CZ$.

\subsection{Formula for the variation of the $\db$-operator}\label{lift}
In order to apply the framework of \S\ref{gq}, we wish to produce a 
family of 
$\db$-operators on a fixed trivialised bundle over $\CP^1$, the family 
being  parametrized
by space-time. We also want the variation with respect to the space-time
coordinates to be given by meromorphic gauge transformations or
diffeomorphisms of $\CP^1$.  This can be done naturally in terms of
the twistor data if a symmetry has been imposed.

Fix a holomorphic symmetry generated by $X$ on $\C^3$, an
$X$-invariant solution to the Bogomolny equations, and corresponding
$\tX$-invariant bundle $E$ on $\CZ$.  Let
$\{\C^2\times\CZ,\db_\alpha\}$ be a Dolbeault representation of $E$
such that $\alpha$ has the canonical form $\alpha = a\,d\bl$ discussed
in \S\ref{wcx}. When pulled back to $\CF$ we obtain a
$\db$-operators $\db_\alpha$ 
on the bundle $\C^2\times\CP^1$ parametrized by
$(v,x,t)\in \C^3$.  

In order to see that the variation of $\db_\alpha$ with respect to the
coordinates on $\C^3$ can be represented in terms of gauge
transformations and diffeomorphisms, note first that $V_0$ and $V_1$
and the Lie derivative operator $\CL_{\hX}$ commute with $\db_\alpha$.
Identifying $\CF$ with $\C^3\times\CP^1$ we have $\hX = X + V$. For
generic $\lambda$, $(X, V_0,V_1)$ span $\C^3$, so a given holomorphic
vector field $Y$ on $\C^3$ can be expressed as
$$
Y = f_0\,V_0 + f_1\,V_1 + h\,X,
$$
where $f_0$, $f_1$ and $h$ are meromorphic in $\lambda$.  Thus
$$
\CL_Y\db_\alpha = \CL_{hX}\db_\alpha = (\CL_{h\hX} -
\CL_{hV})\db_\alpha - [h\Phi,\db_\alpha] 
= - \CL_{hV}\db_\alpha + \db_\alpha (h\Phi),
$$
so that 
\be\label{evolution}
\CL_Y\alpha = - \CL_{hV} \, \alpha + \db_\alpha (h\Phi)\, .
\end{equation}
Thus the derivative of $\db_\alpha$ along $Y$ is given by a
combination of Lie dragging along a meromorphic vector field on
$\CP^1$ and by the action of an infinitesimal meromorphic gauge
transformation.  This gives a natural lift of the action of the Lie
algebra of holomorphic vector fields on $\C^3$ to act on the space of
$\db$-operators on the product bundle $\C^2\times \CP^1$ by a
combination of meromorphic gauge transformations and diffeomorphisms
of $\CP^1$.

\subsection{Formula for the variation of the patching function}
\label{liftred}
One has a similar story in terms of the pullback of the Cech
description.  We start with $P$, a patching function with respect to a
covering by sets $\{ U_+, U_-\}$ for a holomorphic vector bundle with
symmetry on $\CZ$.  The covering can be chosen so that the symmetry
has no fixed points on $U_+$ and so the frames for $E$ on $U_\pm$ can
be chosen so that on $U_+$ is invariant.  Thus on $U_+\cap U_-$ we
will have $\tX P=\phi_- P $ with $\phi_-$ holomorphic on $U_-$.  The
pullback, $\hP$ of $P$ to $\CF$ therefore satisfies $V_0\hP=V_1\hP=0$
and $\hX P= (X+V)\hP=\phi_-\hP$.  So, as before, we can express the
derivative of $\hP$ along a vector field $Y$ on $\C^3$, using $Y=
f_0V_0+f_1V_1+X$ and $\hX=X+V$, as follows
$$
Y\hP=(f_0V_0+f_1V_1+hX)\hP=h(\hX -V)\hP=h(-V\hP +\phi_-\hP) \, .
$$ 
Thus space-time translation corresponds to an action on $\hP$ by
diffeomorphisms and  left multiplication.

\setcounter{equation}{0} 
\section{Definition and formula for the $\tau$-function}

The $\tau$-function was defined in \S\ref{deftau} on an orbit of a
submanifold of the group of gauge transformations to be, in effect,
the determinant of the $\db$ operator on each of the $\CP^1$'s of the
orbit of a given $\db$-operator under that submanifold.  In
\S\ref{lift} the twistor theory was reformulated so that space-time
$\CM$ emerged as the parameter space of a family of holomorphic
structures on a given bundle over $\CP^1$.  Thus $\CM$ can be
naturally embedded into the phase spaces of \S\ref{gq} or \S\ref{red}
and the prequantum bundle $\Det$ can be restricted to it.  For
$\sigma$ to yield a function on space-time, we must trivialize $\Det$
over $\CM$, as in \S\ref{deftau}, by using an invariant trivialization
according to the geometric quantization actions of gauge
transformations and diffeomorphisms given in the previous sections.

\subsection{Dolbeault presentation}
In order to apply the framework of \S\ref{gq} we work locally in
$\C^3$ and decompose $\CP^1=D_+\cup D_-$ so that the poles of $h$ lie
in $D_-\times U$ for the region in $U\subset\C^3$ under consideration.
We also restrict our choice of $\db$-operator so that the support of
$q^*\alpha$ lies in $D_+\times U$.

The tau function $\tau(v,x,t)$ can now be defined as in \S\ref{deftau}
to be the Quillen determinant of $\db_\alpha$.  Using the
infinitesimal version \eqref{deftau1}
we have, for $Y$ a vector field on space-time,
$$
Y\hook \rd \log \tau=\frac{\CL_Y \sigma}{\sigma}
$$
where $\CL_Y$ is as defined in \S\ref{gq}.  It is straightforward to
see that, since the complex gauge transformations and diffeomorphisms
in question extend holomorphically over $D_+$, the central extension
vanishes on the submanifold of $\CG^c$ so that this formula does
indeed define the $\tau$-function up to constants as desired. 
We can make this into an explicit formula for the variation in the
$\tau$-function by applying Proposition \ref{lifg} for the part
associated to the meromorphic gauge transformation $h\Phi$ and formula
\eqref{basic3} for the part arising from the meromorphic
diffeomorphism (our situation satisfies the conditions required for the
simplification leading to that formula) with $\xi^{(1,0)}=hV$.
We obtain
\be\label{tau}
\d\log\tau=-\ip\oint h\tr(\half (Vg)g^{-1}(\d g)g^{-1} +  \Phi(\d g)g^{-1})
\end{equation}
This is equivalent to Proposition 6 (or more directly equation (14)) of
Mason, Singer \& Woodhouse (2000).

\subsection{Cech presentation}
Using equations \eqref{basic2} and \eqref{basic4} these actions can be
lifted to the determinant line bundle and, as in the previous
subsections, we define the $\tau$-function to be the determinant
expressed in an invariant frame.  The formula for its variation along
$Y$ then follows from equations \eqref{basic2} and \eqref{basic4}.
Using the fact that $hV$ is holomorphic on $D_+$ so that the term in
$g_+$ integrates to zero by Cauchy's theorem, This gives
\be\label{tauC}
Y\hook\d\log\tau=-\ip\oint h \tr\left((\half (Vg)g^{-1} +  \Phi)(\d
  g)g^{-1}\right) 
\end{equation}
in which, although formally identical to \eqref{tau} above, the terms
have the different interpretations as in \eqref{basic2} and \eqref{basic4}.

In both cases then, we obtain the same formula for the $\tau$-function
in terms of the pullback to $\CF$ of a holomorphic trivialization of
$E$ on $U_-$ from twistor space.

\setcounter{equation}{0}
\section{Examples}
In this section we apply the above theory to show that the
tau-function for the Ernst equation and the KdV equations according
to the above definitions are given by the standard formulae (see
eg.\ Segal \& Wilson (1985) for KdV and Breitenlohner \& Maison 
(1986) for Ernst).  In each
example we first describe enough of the special features that arise in
the Ward construction to proceed to the calculation of $\tau$ in terms
of the space-time fields using the above theory.  The fact we need 
to do the calculation is that the $P_\pm$ and $g_\pm$ are solutions to
the Lax pair operators.
We give two examples, one in which $hV=0$ and one for which $g\Phi=0$.

\subsection{The KdV equations} For the KdV reduction we have the
symmetry $X=\p/\p v$ on space-time which lifts horizontally to the
spin bundle $\CF$ and descends to $\tX= \p/\p\mu$ on twistor space.
The Bogomolny equations with Lax Pair \eqref{lp1} reduce to the KdV
equation under the further assumption that $B$ is nilpotent and $\tr
(AB)=1$ in which case there exists an invariant gauge in which $D=0$
and
$$
L_0=\p_x + \begin{pmatrix} q& -1\\ p&-q\end{pmatrix}
-\lambda\left(\p_v + 
\begin{pmatrix} 0&0\\ 1&0\end{pmatrix}\right) \, , \qquad L_1=
\p_t +C -\lambda\p_x
$$
The consistency conditions for the Lax pair determine $p$ and $C$ in
terms of $q$ and the condition that $u=\p_xq$ satisfies the KdV
equation:
$$
4\p_t u - \p_x^3u  - 6 u\p_xu=0\, .
$$  

We can make a vector field, $Y=y_0\p_t+y_1\p_x+y_2\p_v$ say, act on
quantities pulled back from twistor space, as in \S\ref{impsymm} as
follows.  Since the lift $\hX$ of $X=\p_v$ to $\CF$ is horizontal
(i.e.\ the $V$ of \S\ref{impsymm} is zero) the distribution $\{\hX,
V_0, V_1\}$ determined by the symmetry and the twistor distribution is
equivalent to the horizontal (over space-time) distribution
$\{\p_v,\p_x,\p_t\}$ on $\CF$.  The leaves are $\lambda=$ constant.
The action of $L_0, L_1$ and $\CL_{\p_v}$ determine a flat connection
on these leaves which is singular at $\lambda=\infty$.  In the
notation of \S\ref{lift} a $Y$ can be chosen to be a symmetry vector
field $Y=a\p_t+b\p_x+c\p_v$.

On twistor space we can choose a Cech presentation based on a framing
on the complement of $\lambda=\infty$ such that
$\CL_{\p_\mu}=\p/\p\mu$ and a framing near infinity so that the lift
of the symmetry $\p/\p \mu$ to the bundle has the form
\begin{equation}
  \label{normform}
\CL_{\p/\p\mu}=\frac{\p}{\p\mu} + \Phi\, , \quad \mbox{ where } \quad
\Phi=  \Phi(\mu,\lambda)=\begin{pmatrix}0 &1/\lambda\\
  1&0\end{pmatrix} + O(\frac{1}{\lambda^2})   
\end{equation}
If the patching function relating the two framings is $P(\mu,\lambda)$
then the symmetry condition is 
$$
\frac{\p P}{\p \mu}+\Phi P=0\, ,
$$
and if $P$ is pulled back to $\CF$, this equation implies
$$
\p_x P+ \lambda \Phi P=0\, , \quad \mbox{ and } \quad \p_t P
+\lambda^2 \Phi P = 0 \, .
$$
The simplest special case is where
$$
P(\mu,\lambda)= \exp\left(-\mu\begin{pmatrix}0 &1/\lambda\\
1&0\end{pmatrix}\right) P(0,\lambda) \, , \quad
\mbox{ 
so that } \quad \Phi= \Phi(\mu,\lambda)=\begin{pmatrix}0 &1/\lambda\\
1&0\end{pmatrix}
$$
exactly.  These bundles correspond to the solutions that Segal and Wilson work
with.  It is not always possible to obtain this normal form except to
finite order in $1/\lambda$.  We will only assume below that it has
this form up to $O(1/\lambda^2)$.

Alternatively one can choose a Dolbeault representation based on a
smooth frame for $E$ which is holomorphic up the fibres of minitwistor
space over $\CP^1$ which agrees with the above framings for
$|\lambda|\geq 1$ but which is not holomorphic for $|\lambda|\leq 1$.
In this frame we will also have $\CL_{\p_\mu}=\p/\p\mu -\Phi$ where
$\Phi$ has the above form for $|\lambda|\geq 1$ but is no longer
explicitly holomorphic for $|\lambda|\leq 1$.  The condition that the
$\db$-operator be invariant implies, with $\bar{\p}_\alpha=\bar{\p}_0
+\alpha$ that
$$
\frac{\p\alpha}{\p\mu}=-\bar{\p}_\alpha\Phi \,  , \quad \mbox{ so that
on $\CF$,}
\quad \p_x\alpha =-\lambda \bar{\p}_\alpha \Phi\, , \mbox{ and } \quad
\p_t\alpha =-\lambda^2 \bar{\p}_\alpha \Phi.
$$
Therefore $\p_x\log\tau$ and $\p_t\log\tau$ are obtained by 
putting $u$ equal to  $\lambda\Phi$ and $\lambda^2\Phi$ (respectively)
in (\ref{basic1}) or (\ref{desc}). In particular,
$$
\p_x\log \tau = \frac{i}{2\pi}\oint \tr \left( (\d
  P_-)P_-^{-1}\lambda \Phi \right)
$$
All the terms in this contour integral are holomorphic in $D_-$,
except the simple pole associated to $\lambda$, so
that it reduces to a residue at $\lambda=\infty$.  Expanding
$P_-=\sum_{i=0}^\infty P_-^i/\lambda^i$ and $\Phi=\sum_{i=0}^\infty
\Phi_i/\lambda^i$ we obtain
$$
\p_x\log \tau = \tr \left( P_-^1(P^0_-)^{-1}\Phi_0\right)
$$

By the definition of the action of the symmetry, we have $\p_v
P_-=-\Phi P_-$. We shall assume that our frame for $E$ has been chosen
so that, to $O(1/\lambda^2)$, $\Phi$ has the normal form given in
equation (\ref{normform}).  Furthermore $P_-$ satisfies the Lax system
(on the right according to our conventions), so we find
$$
\p_vP_-=-P_-\left(\begin{pmatrix} 0&0\\ 1&0\end{pmatrix} - 1/\lambda
\begin{pmatrix} q& -1\\ p&-q\end{pmatrix}\right) + O(1/\lambda^2)
$$ 
so expanding these two equations, we find
$$
\p_vP_-^0=-\Phi_0P^0_-=-P_-^0\begin{pmatrix} 0&0\\ 1&0\end{pmatrix} 
\, , \quad \p_vP_-^1=-\Phi_0P^1_- - \Phi_1P^0_-=-P^1_- \begin{pmatrix}
  0&0\\ 1&0\end{pmatrix}  +P^0_- \begin{pmatrix} q& -1\\
  p&-q\end{pmatrix}  \, . \quad 
$$
Sorting through the equations, we find 
$$
\p_x\log \tau= q
$$
which yields the standard relation between the $\tau$-function and the
solution $u=\p_xq$ to the KdV equation.

\subsection{The Ernst equations}
  The Ernst equations describe vacuum space-times in general
  relativity with two (generic) commuting symmetries that are
  orthogonal to a foliation by 2-surfaces.  We consider a Lorentzian
  metric on $\R^4$ with coordinates $(x^i, r,z)$ 
$$
\d s^2= J_{ij}\d x^i \d x^j - \Omega^2 (\d r^2 + \d z^2)
$$
where $J_{ij}=J_{ij}(r,z)$  is a symmetric $2\times 2$ matrix with
$\det (J)=-r^2$.  The
Einstein vacuum equations on this metric reduce to
$$
\frac{1}{r}\p_r(rJ^{-1}\p_r J) + \p_z(J^{-1}\p_zJ) =0 \, , \quad \mbox{ and
} \quad \frac{\p \log (r\Omega^2)}{\p w} = \frac{i r}{2} \tr \left
( J^{-1}\frac{\p J}{\p w} \right)^2  \quad \mbox{ where } \quad w=z+ir.
$$
The first of these is the reduction of the Bogomolny equations with
a rotational and ${\mathbb Z}_2$ symmetry in a potential form (Ward
1983).  The second determines the conformal factor in terms of $J$ and
will be seen to express $\Omega$ as the tau function associated to
this system according to the above theory (Breitenlohner \& Maison 1986).

We now wish to introduce enough of the twistor correspondence in order
to calculate the expression for $\tau$ above.
We use coordinates $(r,\theta, z)$ on $\R^3$ such that the minitwistor
correspondence becomes
$$
\mu= re^{i\theta} + \lambda z + \lambda^2 re^{-i\theta} \, .
$$
The rotational symmetry $\p/\p\theta $ lifts to give $\p_\theta + i
\lambda\p_\lambda$ on the spin bundle $\CF$ and descends to give
$V=i(\mu\p_\mu +\lambda\p_\lambda)$ on twistor space.

In order to eliminate $\theta$ from the formulae, introduce the
invariant coordinate on $\CF$, $\zeta=e^{-i\theta}\lambda$, and
project to the quotient $\tilde{\CF}$ of $\CF$ by $\p_\theta + i
\lambda\p_\lambda$ with coordinates $(r,z,\zeta)$ (this is equivalent
to expressing the Lax pair in an invariant spin frame).  The Lax pair
becomes
$$
L_0=\p_z - \zeta ( \p_r -\frac{\zeta}{r}\p_\zeta) + J^{-1}\p_z J\, ,
\quad L_1 = \p_r + \frac{\zeta}{r}\p_\zeta + \zeta\p_z + J^{-1}\p_rJ
$$
on $\tilde{\CF}$ and their consistency is equivalent to the first of
the Einstein vacuum equations given above.  These can be reformulated to give
the operators
$$
\p_{\bar{w}} +\frac{1}{2ir(\zeta -i)}\left( (\zeta + i)\zeta\p_\zeta
  +2rJ^{-1}\p_{\bar{w}}J\right) \, , \qquad
\p_{w} +\frac{i}{2r(\zeta + i)}\left( (\zeta - i)\zeta\p_\zeta
  +2iJ^{-1}\p_{w}J\right) \, .
$$
The vector field part of these operators give the form for the map
from vector fields on $\R^3$ to meromorphic vector fields on $\CP^1$
described in \S\ref{impsymm}.  Note that the first operator leads to a
vector field on $\CP^1$ with a pole at $\zeta=i$ and the second at
$\zeta=-i$.  We use a covering of the Riemann sphere so that
$\zeta=\pm i$ are contained in $D_-$.  

The formulae for 
$\d\log\tau$ that are relevant in this case are those for the action
of holomorphic vector fields.
There is no
contribution from $P_+$ (or $g_+$ in the second formulation) as $v$
and $P_+$ are holomorphic over $D_+$.  However there is from $P_-$,
but since the only singularity in $D_-$ is that associated to the
simple pole in $v$, the integral reduces to a residue calculation. 

We therefore find from (\ref{basic3}) or (\ref{basic2}),
$$
\p_{\bar{w}} \log \tau = 
\frac{i}{4\pi} 2\pi i\, \mathrm{Res} 
\{ \frac{(\zeta+i) \zeta}{2ir(\zeta - i)}
\tr  (P_-^{-1}\p P_-)^2 , \zeta=i\} = \frac{1}{2ir}\tr(P_-^{-1}\p_\zeta P_-)^2
|_{\zeta= i}\, .
$$
This last term can be evaluated using the fact that $P_-$ satisfies
the Lax pair so that, at $\zeta=i$
$$
\p_\zeta P_- =r J^{-1}\p_{\bar{w}}J \, P_-
$$
thus giving
$$
\p_{\bar{w}} \log \tau = -\frac{ir}{2} \tr (J^{-1}\p_{\bar{w}}J )^2\, .
$$
We similarly obtain
$$
\p_{w} \log \tau = \frac{ir}{2} \tr (J^{-1}\p_{w}J )^2\, .
$$

We see from this that $\tau$ is equal to $1/r^2\Omega$ up to a
multiplicative constant giving an interpretation to this second part
of the vacuum Einstein equations.

\section{Conclusion and further questions}

In this paper we have described the interrelations that exist between
the twistor description of integrable systems, determinants of
$\db$-operators, and 2-dimensional QFT.

There remain a number of interesting questions.  Can a similar
formulation be found for reductions of the hyper-K\"ahler equations
and their integrable generalizations?  Is it possible to connect the
field theories described above to the topological field theories
involved in the  definition of quantum cohomologies whose partition
functions are also described by the above $\tau$-functions?  Is there
a meaningful way to drop the symmetry assumption on the Bogomolny
equations (or indeed the self-dual Yang-Mills equations) and still
obtain a $\tau$-function description?

\subsection*{Acknowledgements}  We would like to acknowledge useful
discussions with Grame Segal.  MAS would like to thank the EPSRC for
an Advanced Fellowship during which some of this paper was written,
and LJM would like to acknowledge support under Nato Collaborative
Research Grant number 950300. The authors are members of the EDGE
European Network which is funded by the EU (Agreement
No. HPRN-2000-00101).

\section*{References}

\setlength{\parindent}{0in}
\setlength{\parskip}{0.15in}
Atiyah, M. and Bott, R. (1982). The Yang--Mills equations over
Riemann surfaces. {\em Phil. Trans. R. Soc. Lond.} {\bf A308}, 523--615.

Breitenlohner, P., and Maison, D. (1986). On the Geroch group.
{\em Ann.\ Inst.\ H.\ Poincar\'e Phys.\ Theor.} {\bf 46} (1987), 215--46.

Donaldson, S. (1992). Boundary value problems for Yang--Mills
fields. {\em J. Geom. and Phys.} {\bf 8} 89--122.

Donaldson, S. and Kronheimer, P. (1990). The geometry of
4-manifolds. OUP Oxford.

Drinfel'd, V. G., and Sokolov, V. V. (1985). Lie algebras and equations of 
Korteweg--de Vries type.

Ernst, F. (1968): `New formulation of the axially symmetric gravitational 
field problem', {\em Phys.\ Rev.} {\bf 167}, 1175--8. 

Felder, G.,  Gawedzki, K. and  Kupiainen, A. (1988a).
`Spectra of Wess-Zumino-Witten models with arbitrary simple groups',
Commun. Math. Phys.  {\bf 117}, 127.

Felder, G.,  Gawedzki, K. and  Kupiainen, A. (1988b).
`The spectrum of Wess-Zumino-Witten models',
 Nucl. Phys.  {\bf B299}, 355.

Gawedzki, K. and  Kupiainen, A. (1988).
`G/H conformal field theory from gauged WZW model',
Phys. Lett., {\bf B215}, 119.

Gawedzki, K. and  Kupiainen, A. (1989).
`Coset construction from functional integrals',
Nucl. Phys., {\bf B320}, 625,

Gepner, D.  and Witten, E. (1986). 
String theory on group manifolds, Nucl. Phys. {\bf B278}, 493--549.

Guillemin V. and Sternberg, S. (1982) Geometric quantization and
multiplicities of group representations. {\em Invent. Math.} {\bf 67},
515--538.

Hitchin, N. J. (1995). Twistor spaces, Einstein metrics and isomonodromic 
deformations.  {\em J.\ Diff.\ Geom.} {\bf 42}, 30--112.

Jimbo, M., Miwa, T., and Ueno, K. (1981a). Monodromy preserving deformations of 
linear ordinary differential equations with rational coefficients. I. General 
theory and $\tau$-functions. {\em Physica} {\bf 2D}, 306--52.

Jimbo, M., and Miwa, T. (1981b). Monodromy preserving deformations of 
linear ordinary differential equations with rational coefficients. II. 
{\em Physica} {\bf 2D}, 407--48.

Jimbo, M., and Miwa, T. (1981c). Monodromy preserving deformations of 
linear ordinary differential equations with rational coefficients. III. 
{\em Physica} {\bf 4D}, 26--46.

Krichever, I. M. (1994). The $\tau$-function of the universal Whitham 
hierarchy, matrix models and topological field theories. {\em Commun.\ Pure 
Appl.\ Math.} {\bf 47}, 437--75.

Malgrange, B. (1982). Sur les d\'eformations isomonodromiques. {\em 
S\'eminaire de l'Ecole Norm.\ Sup.}, IV, 401--26.

Mason, L. J., (1995).  Twistor theory of the KP equations. in Twistor
theory, ed. S.A.Huggett, Marcel Dekker.

Mason, L. J., and Singer, M. A. (1994). The twistor theory of equations of KdV 
type. I. {Commun.\  Math.\ Phys.} {\bf 166}, 191--218.

Mason, L. J., Singer, M. A. and Woodhouse, N. M. J. (2000). Tau
functions and the twistor theory of integrable systems. Journal of
Geometry and Physics, {\bf 32} (4), 397-430.

Mason, L. J., and Sparling, G. A. J. (1989). Nonlinear Schr\"odinger and 
Korteweg de Vries are reductions of self-dual Yang-Mills. {Phys.\ Lett.} {\bf 
A137}, 29--33.

Mason, L. J., and Sparling, G. A. J. (1992). Twistor correspondences for the 
soliton hierarchies. {\em J. Geom.\ Phys.} {\bf 8}, 243--71.

Mason, L. J., and Woodhouse, N. M. J. (1988). 
The Geroch group and non-Hausdorff twistor spaces. {\em Nonlinearity} {\bf 1}, 
73--114

Mason, L. J., and Woodhouse, N. M. J. (1996). {\em Integrability, 
self-duality, and twistor theory}, Oxford.

Miwa, T. (1981). Painlev\'e property of the monodromy preserving deformation equations 
and the analyticity of $\tau$ functions. {\em Publ.\ RIMS}, Kyoto, 
{\bf 17}, 703--21.

Palmer, J. (1993). Tau functions for the Dirac operator in the Euclidean plane. {\em 
Pacific J. Math.} {\bf 160}, 259--342.

Pressley, A.,  and Segal, G. (1986). {\em Loop groups}. Oxford University 
Press, Oxford.

Quillen, D. (1985).  Determinants of Cauchy-Riemann operators over Riemann 
surface. {\em Funct.\ anal.\ appl.} {\bf 19}, 37--41.

Reed, M., and Simon, B. (1978). {\em Methods of modern mathematical physics} 
IV. {\em Analysis of operators}. Academic Press, New York.

Sato, M., Miwa, T., and Jimbo, M. (1978). Holonomic quantum fields. I. 
{\em Publ.\ RIMS}, Kyoto, {\bf 14}, 223-67.

Sato, M., Miwa, T., and Jimbo, M. (1979). Holonomic quantum fields. II--IV. 
{\em Publ.\ RIMS}, Kyoto,  {\bf 15}, 201-67, 577-629, 871--972.

Sato, M., Miwa, T., and Jimbo, M. (1980). Holonomic quantum fields. V. {\em 
Publ.\ RIMS}, Kyoto, {\bf 16}, 531--84. 

Segal, G. (1989).  Two-dimensional conformal field theories and
modular functors. In {\em IXth International Congress on Mathematical
  Physics (Swansea 1988)}, p 22--37, Hilger Bristol.  See also: Segal,
G. (1991).  The definition of conformal field theory, (1991) Oxford
preprint (unpublished).

Segal, G., and Wilson, G. (1985). Loop groups and equations of KdV type. {\em 
Pub.\ Math.} {\bf 61}, IHES.

Takasaki, K. (1995). Symmetries and tau function of higher dimensional 
integrable hierarchies. {\em J. Math.\ Phys.} {\bf 36}, 3574--607.

Ward, R. S. (1983). `Stationary axisymmetric space-times: a new approach', 
{\em Gen.\ Rel.\ Grav.} {\bf 15}, 105--9.

Ward, R. S. (1984). The Painlev\'e property for the self-dual gauge field 
equations.  {\em Phys\. Lett} {\bf 102A}, 279--82.

Ward, R. S. (1986). Integrable and solvable systems and relations among them.
{\em Phil.\ Trans.\ R.\ Soc.} {\bf A315}, 451--7.

Ward, R. S. \& Wells, R. O. (1990). {\em Twistor Geometry and Field Theory.}
CUP.

Witten, E. (1984). `Nonabelian bosonization in two dimensions',
Commun. Math. Phys. {\bf 92}, 455--472.

Witten, E. (1988). `Quantum field theory, Grassmanians, and algebraic curves',
Commun. Math. Phys., {\bf 113}, 529-600. 

Witten, L. (1979). `Static axially symmetric solutions of 
self-dual ${\rm SU}(2)$ gauge fields in Euclidean four-dimensional space', 
{\em Phys.\ Rev.} {\bf D19}, 718--20.

Woodhouse, N. M. J. (1992). {\em Geometric quantization}. 2nd edition. Oxford 
University Press, Oxford.
\end{document}